\documentclass[11pt]{article}
\usepackage[a4paper,margin=2.4cm]{geometry}
\usepackage[T1]{fontenc}\usepackage[utf8]{inputenc}\usepackage{lmodern,microtype}
\usepackage{amsmath,amssymb,amsthm,booktabs,tabularx,graphicx,hyperref,xcolor}
\usepackage{tikz}\usetikzlibrary{arrows.meta,positioning}
\usepackage{placeins}
\hypersetup{colorlinks=true,linkcolor=blue!50!black,urlcolor=blue!50!black,citecolor=blue!50!black}
\newtheorem{theorem}{Theorem}\newtheorem{proposition}{Proposition}\newtheorem{remark}{Remark}
\newcommand{\F}{\mathbb F_2}\newcommand{\rank}{\operatorname{rank}}\newcommand{\xor}{\oplus}
\title{Advantage Distillation in Paired-State Quantum Key Distribution}
\author{Luis Adri\'an Lizama-P\'erez$^{1}$ and Jos\'e Mauricio L\'opez-Romero$^{2}$\\[.5em]
\small $^{1}$Universidad Aut\'onoma Metropolitana, Unidad Lerma, Mexico\\
\small $^{2}$Cinvestav Quer\'etaro, Mexico\\
\small Correspondence: \texttt{l.lizama@ler.uam.mx}}
\date{November 2026}
\begin{document}\maketitle
\begin{abstract}
We investigate advantage distillation for paired-state Loop-Back quantum key distribution, with a fixed optical station serving a lightweight portable terminal. Alice prepares two polarization-encoded pulses, Bob selects and applies a polarization filter to each pulse, and Alice infers his selections from the returned detection pattern. The proposed advantage-distillation step selects consistent blocks of three two-bit records using two independent public parity checks on a minimal connected graph. The closing edge of a parity polygon is shown to be redundant: any spanning tree of the same vertices has the same acceptance set. We derive exact acceptance probabilities, residual-error laws, and retention rates. Under independent bit errors, the three-record filter reduces the leading error term from $e$ to $e^2$, while its cost in discarded records and public information is accounted for explicitly. A modular security theorem bounds the additional guessing advantage available to an adversary after syndrome disclosure and block acceptance, including arbitrary quantum side information. We also evaluate a specified single-photon passive-tap optical channel, obtaining a positive conditional entropy bound and exact accepted-block security quantities. This parity-based advantage-distillation step detects discrepancies by postselection; it does not correct individual errors and is intended to precede conventional reconciliation. The architecture suggests two further directions: hybrid filtering and reconciliation, and key establishment between portable users accessing a shared quantum station.
\end{abstract}

\section{Introduction}
A mobile user need not carry a complete quantum transmitter and receiver to take part in quantum key distribution. In the Loop-Back architecture, a fixed station prepares the optical signals and measures their return, while a portable terminal performs a comparatively simple optical operation. This division of hardware suggests a practical route toward quantum communication infrastructure shared by many users \cite{loop,recent}. In the present work we examine one such link: Alice operates the station, and Bob selects polarization filters that encode two bits per valid pair of pulses.

The optical exchange and the subsequent classical processing address different problems. Alice can reconstruct Bob's filter settings in the ideal model, including settings inferred from a missing return. In a real channel, however, a non-click can also arise from loss. The resulting discrepancies must be detected or reconciled before a common secret key can be extracted. Existing QKD systems frequently use error-correcting procedures such as Winnow, Cascade, and low-density parity-check (LDPC) decoding \cite{winnow,cascade,ldpc}. The broader principle of distilling secret correlations through public discussion goes back to Maurer's foundational treatment \cite{maurer1993}. Here we study a complementary parity-based advantage-distillation step: comparing public parities and discarding blocks that fail the test. This operation reduces errors by selection rather than changing the values of incorrect bits; whether it increases the final secret-key rate depends on the information available to Eve and the cost of public processing.

Earlier work on non-orthogonal state pairs, binary frames, reverse reconciliation, and Loop-Back provides the context for this construction \cite{flows,nack,frames,three,conj,reverse,loop}. Our emphasis is on determining how few independent checks are needed and how much can be established from them. The closed polygon that motivated the method has a redundant final edge. A connected path on three two-bit records uses only two public checks and accepts exactly the same error patterns as the corresponding triangle. The acceptance condition can be characterized algebraically for any connected graph, and the residual error distribution can be calculated exactly under the stated noise model.

Two qualifications shape the analysis rather than diminishing its purpose. First, a lower error rate does not alone imply a higher final secret-key rate: discarded blocks, disclosed checks, and subsequent reconciliation all carry costs. Second, the uncertainty of an adversary about Bob's bits must be established from an optical attack model; it cannot be inferred solely from the observed error rate. We therefore give a security reduction valid for arbitrary quantum side information when a pre-filter guessing-probability bound is available, and derive such a bound for a specified passive-tap channel. The general theory and its physical example are kept separate so that stronger optical models can be incorporated later.

The paper proceeds from the preparation and filtering of the optical pulses to the minimal graph, its exact error laws, and its security analysis. We close with two lines of work that follow naturally from the results: combining the filter with established reconciliation codes, and extending the station to serve two portable users who wish to share a key that remains private from the station.

\section{Optical generation of two-bit records}\label{sec:physical}
\subsection{Preparation, polarization filtering, and bit inference}
Alice selects independent uniform bits $a_X,a_Z$ and prepares
\begin{equation}
 |a_X\rangle_X\otimes |a_Z\rangle_Z.
\end{equation}
The pulses occupy distinct, identified time slots. To make the basis-sifting probability explicit, we use the following implementation convention: Alice randomly orders the bases as $(X,Z)$ or $(Z,X)$, and Bob independently selects one of the same two orders. For each slot Bob independently chooses a uniform polarizer value $b\in\{0,1\}$ in his selected basis. He returns the transmitted field, including vacuum when his polarizer blocks the photon. Alice measures the return in her preparation bases. Bob announces his basis order, but keeps the two polarizer values private. They retain the orders that agree, with probability $p_{\rm sift}=1/2$, and reorder each retained record canonically as $(X,Z)$.

The random-order convention fixes $p_{\rm sift}=1/2$ for the implementation studied here. If Bob instead independently chooses a basis for each slot and both must agree, the sifting probability is $1/4$. None of the conditional filter laws below depends on which convention is used, but per-emission accounting does.

A basis is $X$ or $Z$; a polarizer value is $0$ or $1$ within that basis. We use basis labels exclusively for $X$ and $Z$, and binary values for the corresponding polarizer settings. On a matched slot, a selective ideal polarizer has transmitted amplitude
\begin{equation}
 P_b^{(\theta)}|a\rangle_\theta
 =\delta_{ab}|a\rangle_\theta,\qquad
 P_b^{(\theta)}=|b\rangle_\theta\langle b|,\quad \theta\in\{X,Z\}.
\end{equation}
The missing norm is absorption into the blocked outcome, represented as vacuum on the return path. The filter transmits the selected polarization component and removes the orthogonal component from the return. It should not be confused with a unitary rotation, which transforms the polarization without such selection.

\begin{figure}[htbp]\centering
\begin{tikzpicture}[>=Latex,font=\small,
 station/.style={draw=blue!55!black,fill=blue!3,rounded corners=3pt,align=center,text width=3.7cm,minimum height=1.8cm,inner sep=6pt},
 note/.style={align=center,text width=3.6cm,font=\footnotesize}]
\node[station] (prep) at (0,0) {\textbf{Alice prepares}\\[3pt]$|a_X\rangle_X,\ |a_Z\rangle_Z$\\Two timed pulses};
\node[station] (bob) at (4.9,0) {\textbf{Bob filters}\\[3pt]$P_{b_X}^{(X)},\ P_{b_Z}^{(Z)}$\\Private values $b_X,b_Z$};
\node[station] (measure) at (9.8,0) {\textbf{Alice measures}\\[3pt]$C_X,C_Z\in\{0,1\}$\\Timed return windows};
\draw[->,thick,blue!60!black] (prep.east)--node[above,font=\footnotesize]{outward}(bob.west);
\draw[->,thick,blue!60!black] (bob.east)--node[above,font=\footnotesize]{return}(measure.west);
\node[note] at (0,-1.7) {\textbf{Ideal example}\\$|0\rangle_X,\ |0\rangle_Z$};
\node[note] at (4.9,-1.7) {Bob chooses $(0_X,1_Z)$\\First pulse passes\\Second pulse is blocked};
\node[note] at (9.8,-1.7) {Return: $|0\rangle_X,\ |\mathrm{vac}\rangle$\\Clicks: $(C_X,C_Z)=(1,0)$};
\node[draw=blue!55!black,fill=blue!3,rounded corners=3pt,align=center,text width=12.9cm,inner sep=8pt] (rec) at (4.9,-3.25)
 {\textbf{After matching the basis orders:} $\widetilde b_\theta=a_\theta\oplus(1-C_\theta)$\\[4pt]
 Alice reconstructs $(0\oplus0,\ 0\oplus1)=(0,1)$; Bob's target record is $K_i=(0,1)$.};

\end{tikzpicture}
\caption{The two pulses travel from Alice to Bob and back to Alice. The two Alice boxes represent preparation and detection at the same station. The worked example assumes matching basis orders, single photons and no accidental loss. A non-click in the second assigned window reveals deliberate blocking only under these ideal assumptions.}
\label{fig:roundtrip}
\end{figure}
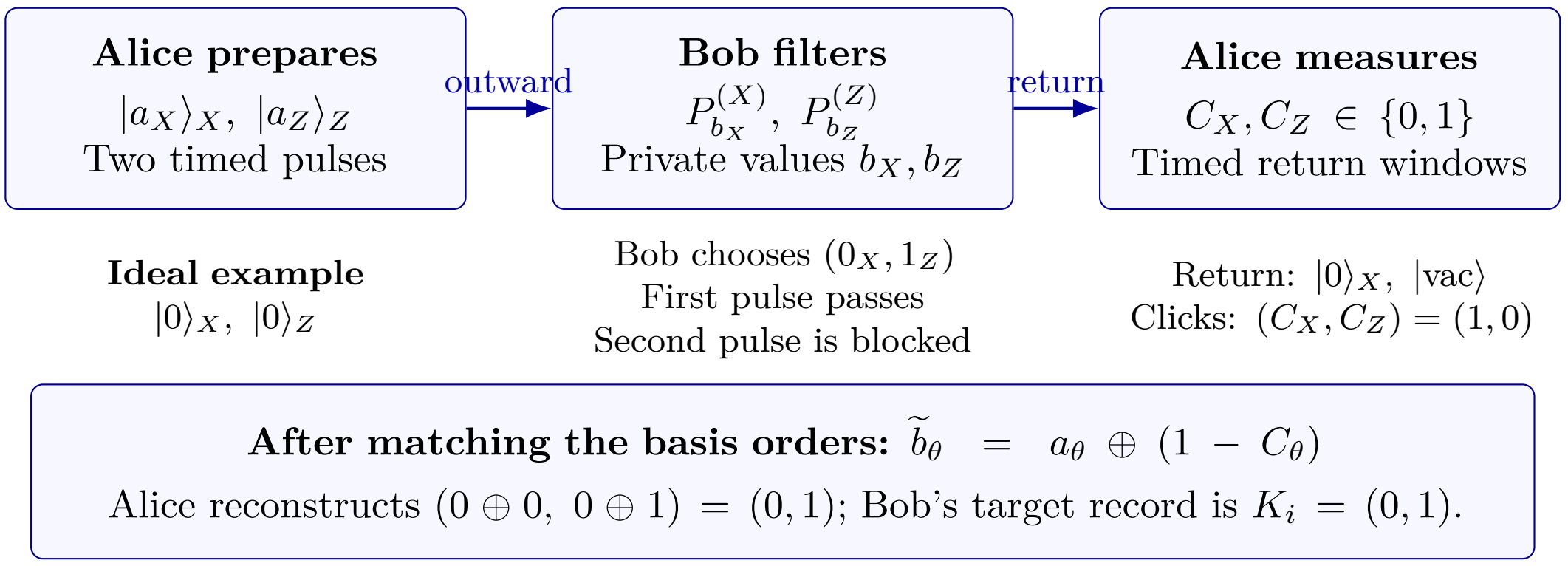

Let $C_\theta=1$ denote a click in the monitored preparation-polarization port and $C_\theta=0$ a non-click in its assigned return window. In the ideal model no opposite-port or multiple-click outcomes occur; any extension with those outcomes must specify how they are classified or discarded. With ideal single photons, no channel loss and perfect detection,
\begin{equation}
 C_\theta=\mathbf1_{a_\theta=b_\theta},\qquad
 \widetilde b_\theta=a_\theta\xor(1-C_\theta)=b_\theta.\label{eq:reconstruction}
\end{equation}
Thus Bob's record is $K_i=(b_{X,i},b_{Z,i})$ and Alice's estimate is $A_i=(\widetilde b_{X,i},\widetilde b_{Z,i})$. Alice's preparation bits are auxiliary private data, not the target raw key.
Figure~\ref{fig:roundtrip} follows one ideal pair from preparation to the reconstructed record. The same rule applies to the four detection patterns in Table~\ref{tab:map}.

\begin{table}[ht]\centering
\caption{Ideal reconstruction for preparation $(a_X,a_Z)=(0,0)$, after matching basis orders. Both non-click windows are retained.}
\begin{tabular}{@{}ccc@{}}\toprule
Alice's clicks $(C_X,C_Z)$ & Bob's polarizer values & Reconstructed record $A_i$\\\midrule
$(1,1)$ & $(0_X,0_Z)$ & $00$\\
$(1,0)$ & $(0_X,1_Z)$ & $01$\\
$(0,1)$ & $(1_X,0_Z)$ & $10$\\
$(0,0)$ & $(1_X,1_Z)$ & $11$\\\bottomrule
\end{tabular}\label{tab:map}
\end{table}
\begin{figure}[htbp]\centering
\begin{tikzpicture}[font=\footnotesize,>=Latex,
 cell/.style={draw=black!45,rounded corners=2pt,minimum width=2.1cm,minimum height=.74cm,align=center},
 header/.style={font=\footnotesize\bfseries},
 pass/.style={circle,fill=blue!65!black,inner sep=2.5pt},
 absent/.style={circle,draw=black!65,thick,inner sep=2.1pt}]
\node[header,anchor=west] at (0,2.13) {Bob's two filters};
\node[header,anchor=west] at (3.02,2.13) {Return $X,Z$};
\node[header,anchor=west] at (6.05,2.13) {Alice's bits};
\foreach \y/\bx/\bz/\cx/\cz/\bits in {1.55/0/0/1/1/00,.55/0/1/1/0/01,-.45/1/0/0/1/10,-1.45/1/1/0/0/11}{
  \node[cell,anchor=west] at (0,\y) {$(\bx_X,\bz_Z)$};
  \draw[->,black!55] (2.2,\y)--(2.93,\y);
  \node[draw=black!35,minimum width=2.15cm,minimum height=.74cm,anchor=west] at (3.02,\y) {};
  \ifnum\cx=1 \node[pass] at (3.58,\y) {}; \else \node[absent] at (3.58,\y) {};\fi
  \ifnum\cz=1 \node[pass] at (4.57,\y) {}; \else \node[absent] at (4.57,\y) {};\fi
  \draw[->,black!55] (5.3,\y)--(5.97,\y);
  \node[cell,anchor=west] at (6.05,\y) {$\bits$};
}
\node[anchor=west,font=\scriptsize,text=black!70] at (0,-2.12) {Alice prepared $(|0\rangle_X,|0\rangle_Z)$ in every row; filled $=$ click, open $=$ no click.};
\end{tikzpicture}
\caption{Four ideal outcomes for the same prepared pair $(|0\rangle_X,|0\rangle_Z)$, with matching basis orders. A filled dot denotes a registered return click and an open circle denotes no return click in the assigned time slot. Bob selects the two filter values; Alice reconstructs them from the click pattern. The no-click events are information-bearing only under the stated ideal optical model.}
\label{fig:idealoutcomes}
\end{figure}
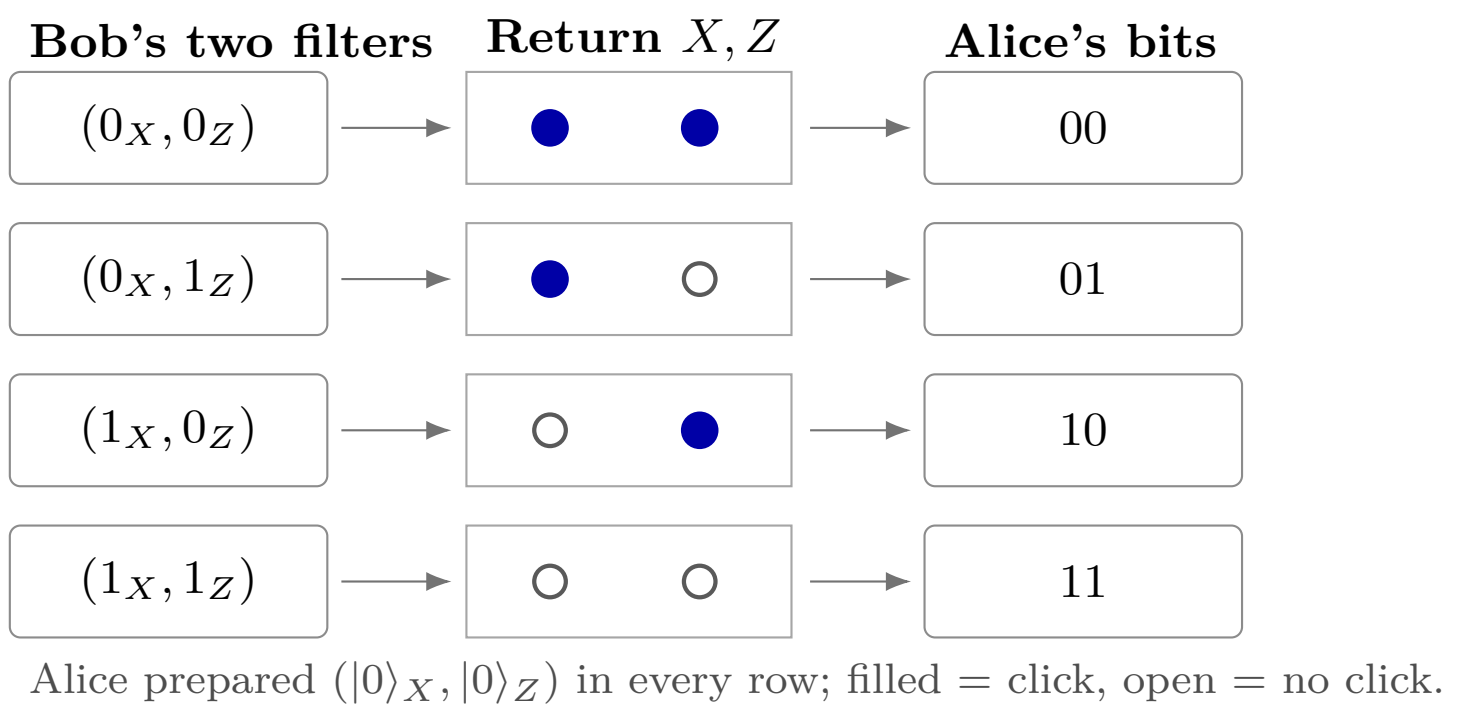

\begin{figure}[htbp]\centering
\resizebox{\linewidth}{!}{
\begin{tikzpicture}[x=1cm,y=1cm,>=Latex,font=\footnotesize,
  sphere/.style={circle,draw=cyan!40!black,fill=cyan!11,minimum size=.91cm,inner sep=0pt,line width=.55pt},
  flow/.style={-{Latex[length=2mm]},black!70,line width=.65pt},
  filterbox/.style={draw=black!45,densely dashed,rounded corners=4pt},
  phot/.style={circle,fill=black,minimum size=3.2pt,inner sep=0pt},
  vac/.style={circle,draw=black,line width=.8pt,fill=white,minimum size=5pt,inner sep=0pt}]
\node[anchor=west,font=\bfseries] at (.05,7.27) {(a) Paired pulses prepared by Alice};
\node[font=\bfseries] at (2.5,6.78) {Alice};
\node at (1.72,6.42) {$X$ pulse};\node at (3.28,6.42) {$Z$ pulse};
\node[sphere] (ax) at (1.72,5.81) {};
\node[sphere] (az) at (3.28,5.81) {};
\node[phot] at (2.17,5.81) {};
\node[phot] at (3.28,6.26) {};
\node[anchor=west,font=\scriptsize] at (2.21,5.91) {$0_X$};
\node[anchor=west,font=\scriptsize] at (3.73,6.13) {$0_Z$};
\node at (2.5,5.12) {$\{\lvert0\rangle_X,\lvert0\rangle_Z\}$};
\draw[flow] (4.23,5.81)--(9.75,5.81) node[midway,above] {two time-separated pulses};
\node[font=\bfseries] at (11.4,6.44) {Bob};
\node[draw=cyan!35!black,fill=cyan!7,rounded corners=4pt,minimum width=3.15cm,minimum height=.8cm] at (11.4,5.81) {polarization filters};
\node[anchor=west,font=\bfseries] at (.05,4.49) {(b) Bob's filter choices and ideal return outcomes};
\node[anchor=west,font=\scriptsize] at (.1,3.94) {Alice's input: $\lvert0\rangle_X,\lvert0\rangle_Z$};
\node[font=\bfseries] at (6.2,3.95) {Bob};
\node[font=\bfseries] at (11.0,3.95) {Returned to Alice};
\node at (5.28,3.55) {$X$ filter};\node at (7.0,3.55) {$Z$ filter};
\node at (10.45,3.55) {$X$ pulse};\node at (12.15,3.55) {$Z$ pulse};
\foreach \y/\xb/\zb/\cx/\cz/\bits in {2.95/0/0/1/1/00,1.95/0/1/1/0/01,.95/1/0/0/1/10,-.05/1/1/0/0/11}{
  \node[anchor=west] at (.2,\y) {$(\xb_X,\zb_Z)$};
  \draw[filterbox] (3.99,\y-.43) rectangle (7.86,\y+.43);
  \node[sphere] at (5.28,\y) {};
  \node[sphere] at (7.0,\y) {};
  \ifnum\xb=0
    \draw[-{Latex[length=1.7mm]},line width=.8pt] (5.05,\y)--(5.59,\y);
  \else
    \draw[-{Latex[length=1.7mm]},line width=.8pt] (5.52,\y)--(4.98,\y);
  \fi
  \ifnum\zb=0
    \draw[-{Latex[length=1.7mm]},line width=.8pt] (7.0,\y-.30)--(7.0,\y+.28);
  \else
    \draw[-{Latex[length=1.7mm]},line width=.8pt] (7.0,\y+.30)--(7.0,\y-.28);
  \fi
  \draw[flow] (2.15,\y)--(3.81,\y);
  \draw[flow] (7.97,\y)--(9.38,\y);
  \node[sphere] at (10.45,\y) {};
  \node[sphere] at (12.15,\y) {};
  \ifnum\cx=1
    \node[phot] at (10.9,\y) {};
  \else
    \node[vac] at (10.9,\y) {};
  \fi
  \ifnum\cz=1
    \node[phot] at (12.15,\y+.45) {};
  \else
    \node[vac] at (12.15,\y+.45) {};
  \fi
  \node[anchor=west,font=\scriptsize] at (13.08,\y) {$C_XC_Z=\cx\cz$};
}
\node[phot] at (1.05,-.81) {};
\node[anchor=west,font=\scriptsize] at (1.27,-.81) {transmitted / click};
\node[vac] at (5.14,-.81) {};
\node[anchor=west,font=\scriptsize] at (5.38,-.81) {blocked / no click};
\node[anchor=west,font=\scriptsize,text=black!65] at (9.00,-.81) {Ideal lossless, basis-matched case};
\end{tikzpicture}}
\caption{Paired-pulse preparation and four ideal polarization-filter settings in the matched $X,Z$ order. Alice prepares the example $(|0\rangle_X,|0\rangle_Z)$, and Bob independently chooses the transmitted polarization of each temporal slot. Black marks identify photons transmitted into Alice's return mode; open marks indicate a blocked return (vacuum), not a physical Bloch-sphere state. The resulting click patterns are $(1,1)$, $(1,0)$, $(0,1)$ and $(0,0)$, allowing Alice to reconstruct Bob's two filter values under the ideal lossless model. The drawings are schematic Bloch-sphere projections rather than representations of measured trajectories.}\label{fig:bob_bloch_filter}
\end{figure}
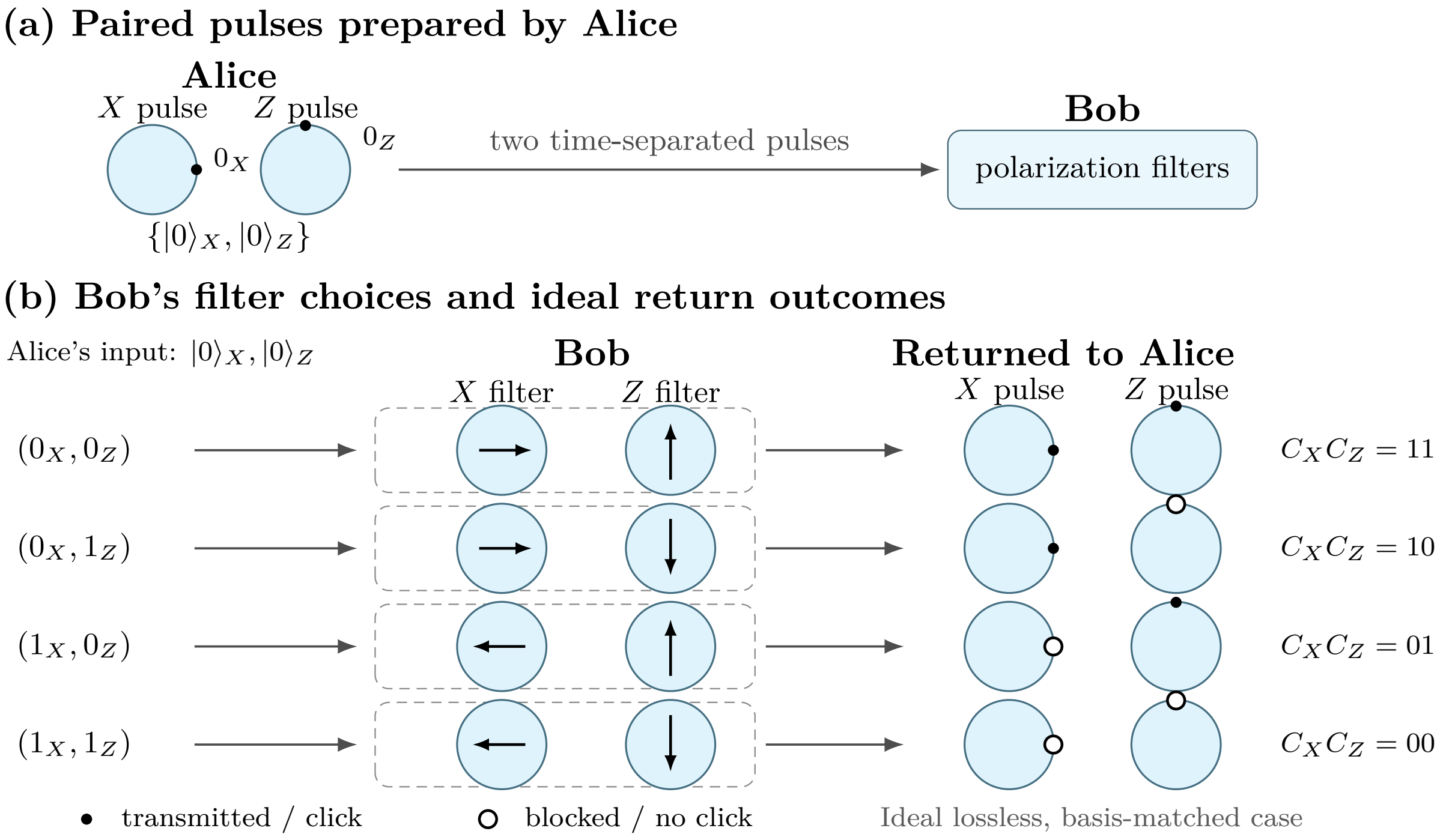

The polarization geometry underlying the matched-basis example is shown in Figure~\ref{fig:bob_bloch_filter}. Its disk projections follow the visual convention of the Bloch-sphere diagrams used in earlier paired-state work. The arrows specify Bob's independently selected polarizer values. A filled mark denotes a transmitted pulse in the ideal return mode; an open mark denotes no returned photon, not another point on the Bloch sphere. Thus the figure emphasizes that Bob applies a \emph{selective filter}, not a state rotation.

\paragraph{Two-bit record representation.}
The same preparation $(0_X,0_Z)$ is used in Figure~\ref{fig:idealoutcomes} solely to isolate Bob's choices. In an actual run Alice randomizes both preparation values and their slot order. A retained pair is indexed as one record, even when neither return window clicks. For example, Bob's filters $(0_X,1_Z)$ produce $(C_X,C_Z)=(1,0)$ and the inferred record $A_i=(0,1)$, which agrees with $K_i=(0,1)$. Four independent retained pairs will supply four vertices in the illustrative cycle below. The figures describe ideal honest-operation events; secrecy against an adversary is analyzed separately.

Each of the four two-bit records is allowed, and its within-record parity can be either zero or one. An event means an indexed emitted pair with two scheduled return windows, not necessarily two detected photons. Timing and emission indexing are essential to distinguish an assigned non-click from the absence of an attempted transmission.

All basis-matched pairs enter the baseline filter, including zero-click pairs. Detection categories are useful for diagnosis but are not silently used to construct or select graph blocks. If category-based selection is introduced, its public messages and conditional error law must be included explicitly.

\subsection{Optical implementation and related experimental architectures}
The selective action assigned to Bob is an idealized polarization-dependent transmission operation, followed by routing of the transmitted component toward Alice. Polarization-selective optics and modulating retroreflectors provide relevant implementation precedents. Rabinovich \emph{et al.} demonstrated a laboratory free-space BB84 link employing modulating retroreflectors and high-extinction multiple-quantum-well modulators \cite{rabinovich2018}. Zhu, Hu and Guo subsequently investigated a retroreflector-based B92 arrangement with polarization splitting and intensity modulation, including the influence of finite extinction and optical misalignment \cite{zhu2022}. These studies establish the availability of related optical functions, but neither implements the present two-slot polarizer-value selection and click/no-click reconstruction rule. Bidirectional QKD using Faraday-mirror compensation provides a further precedent for round-trip propagation and polarization stabilization in fiber systems \cite{muller1997}; its interferometric encoding is distinct from the selective filtering considered here.

For the present protocol, Bob must address two identified temporal slots, select independently the transmitted polarization within the prescribed $X$ and $Z$ bases, and direct only the accepted optical component into the return channel. The orthogonal component must not contribute to Alice's monitored return window. Thus the ideal blocked outcome refers to absence of light in that return mode; absorption within a particular component is not required as a universal physical implementation. Finite extinction may produce unwanted return clicks, whereas insertion loss, propagation attenuation and detector inefficiency may produce missing clicks for nominally transmitted states. Polarization drift, return-path routing, timing accuracy and the optical response to externally injected light would likewise require characterization. The parameters $\alpha$ and $\beta$ below describe the corresponding honest-operation click statistics, but do not replace a component-level implementation or a device-security model. No experimental validation of this specific Bob module is claimed.

\subsection{Detection model and inference errors}
For each retained slot, assume a binary response with
\begin{equation}
 \Pr(C=1\mid a=b)=\alpha,\qquad
 \Pr(C=1\mid a\ne b)=\beta.\label{eq:clickmodel}
\end{equation}
The response is assumed symmetric under simultaneous bit complementation and identical in the two bases. Define $\epsilon=\widetilde b\xor b$. With independent uniform $a,b$, half the inputs have equal values. An equal-value input is erroneous on a non-click, whereas an opposite-value input is erroneous on a click. Consequently,
\begin{equation}
 e=\Pr(\epsilon=1)=\frac{1-\alpha+\beta}{2}.\label{eq:physicalerror}
\end{equation}
For either fixed value of $b$, the same expression holds after averaging over $a$; hence $\epsilon$ is independent of Bob's uniform target bit in this honest model. Independent preparation bits, polarizer bits and response randomness across slots yield independent coordinate errors. The two-coordinate probabilities are then
\begin{equation}
 \Pr(\epsilon_X,\epsilon_Z)=[(1-e)^2,e(1-e),e(1-e),e^2].\label{eq:physicalpair}
\end{equation}
The resulting iid benchmark follows from the specified honest-operation model and does not constrain adversarial disturbances. Common-mode losses, detector memory, imperfect extinction or correlated preparations can violate it.

For polarization-preserving loss only, let
\begin{equation}
 \eta=\eta_{\rightarrow}\eta_{\leftarrow}\eta_{\rm pol}\eta_{\rm det}
\end{equation}
be the probability that an unblocked emitted photon produces a monitored click, including both propagation legs, polarizer insertion transmission and detection. Then $\alpha=\eta$, $\beta=0$, and
\begin{equation}
 e=\frac{1-\eta}{2}.\label{eq:loss}
\end{equation}
The error comes from interpreting accidental loss as deliberate blocking. There is no contradiction with a polarization-preserving channel: the wrong bit is a wrong inference of Bob's operation.

For an additional independent background-click probability $\delta$ per monitored window, with no dead time or afterpulsing,
\begin{equation}
 \alpha=1-(1-\eta)(1-\delta),\quad \beta=\delta,\quad
 e=\frac{1-\eta+\eta\delta}{2}.\label{eq:dark}
\end{equation}
This is a threshold-click model for a single monitored port. It is not a general model of polarization noise or multi-photon illumination. Weak coherent sources require photon-number statistics and a corresponding security treatment; substituting a mean intensity for $\eta$ does not supply that treatment.

\subsection{Error probabilities conditioned on detection events}
Writing $p_C=(\alpha+\beta)/2$, independent slots yield Type-1 (two-click), Type-2 (one-click) and Type-3 (zero-click) probabilities $p_C^2$, $2p_C(1-p_C)$ and $(1-p_C)^2$. Bayes' rule gives
\begin{equation}
 e_{\rm click}=\frac{\beta}{\alpha+\beta},\qquad
 e_{\rm no\ click}=\frac{1-\alpha}{2-\alpha-\beta},\label{eq:categories}
\end{equation}
when the conditioning event has nonzero probability. Conditional on a specified two-slot click pattern, errors remain independent in this model, but the two coordinates generally have different error rates. The probability of any error in a Type-1 pair is $2e_{\rm click}-e_{\rm click}^2$; the probability of two errors is $e_{\rm click}^2$. Neither is generically the unconditional $e^2$.

In the loss-only model, Type-1 pairs are indeed clean, because a click certifies equality in this idealized device model. Type-2 and Type-3 non-click coordinates have error $(1-\eta)/(2-\eta)$. That clean-Type-1 conclusion fails as soon as $\beta>0$ and is not an adversarial certificate. Public category selection must also be accounted for. The baseline retains all three categories so that equation~\eqref{eq:physicalpair} applies without category conditioning.

\FloatBarrier
\section{Minimal connected parity filtering}\label{sec:polygon}
Each retained two-pulse exchange gives Bob a two-bit record $K_i=(b_{X,i},b_{Z,i})$ and Alice an estimate $A_i=(\widetilde b_{X,i},\widetilde b_{Z,i})$. Graph vertices represent complete records, not single photons. Define the local binary parities
\begin{equation}
 p_i^B=b_{X,i}\xor b_{Z,i},\qquad p_i^A=\widetilde b_{X,i}\xor\widetilde b_{Z,i}.
\end{equation}
An edge $\{i,j\}$ tests $s_{ij}=p_i^B\xor p_j^B$ against $u_{ij}=p_i^A\xor p_j^A$. Bob broadcasts the edge syndromes on an authenticated channel; Alice announces which blocks pass. The decision rule is $F=1$ if and only if $s_{ij}=u_{ij}$ for every selected edge. Every failed block is discarded in full; the procedure neither identifies nor corrects individual coordinate errors.

\subsection{Minimal three-record construction}
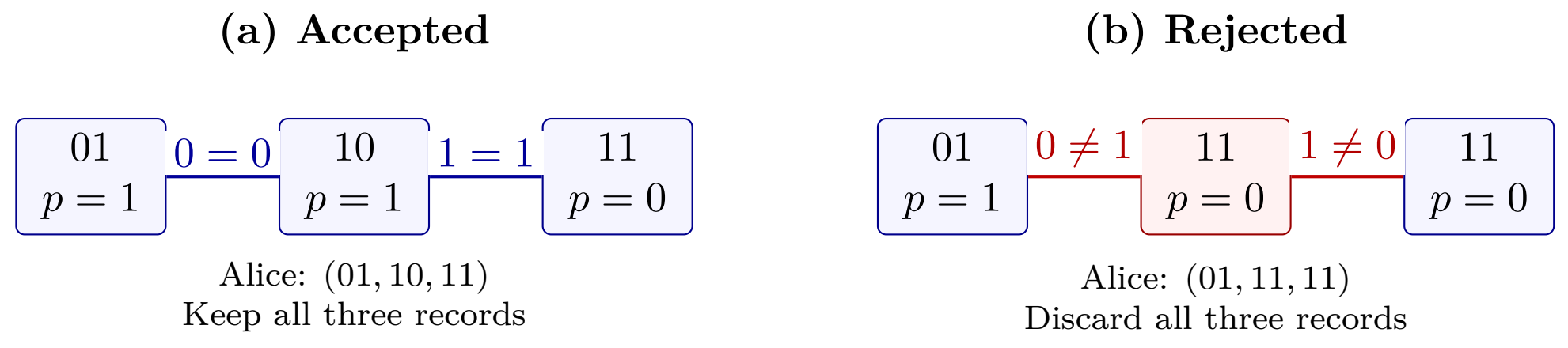
\begin{figure}[htbp]\centering
\begin{tikzpicture}[font=\small,
 box/.style={draw=blue!55!black,fill=blue!4,rounded corners=2pt,minimum width=1.25cm,minimum height=0.75cm,align=center},
 altered/.style={draw=red!60!black,fill=red!5,rounded corners=2pt,minimum width=1.25cm,minimum height=.75cm,align=center}]
\begin{scope}
\node[font=\small\bfseries] at (2.2,1.2) {(a) Accepted};
\node[box] (a1) at (0,0) {$01$\\$p=1$};
\node[box] (a2) at (2.2,0) {$10$\\$p=1$};
\node[box] (a3) at (4.4,0) {$11$\\$p=0$};
\draw[blue!60!black,thick] (a1)--node[above,fill=white,inner sep=2pt]{$0=0$}(a2);
\draw[blue!60!black,thick] (a2)--node[above,fill=white,inner sep=2pt]{$1=1$}(a3);
\node[font=\scriptsize,align=center] at (2.2,-1) {Alice: $(01,10,11)$\\Keep all three records};
\end{scope}
\begin{scope}[xshift=7.2cm]
\node[font=\small\bfseries] at (2.2,1.2) {(b) Rejected};
\node[box] (b1) at (0,0) {$01$\\$p=1$};
\node[altered] (b2) at (2.2,0) {$11$\\$p=0$};
\node[box] (b3) at (4.4,0) {$11$\\$p=0$};
\draw[red!75!black,thick] (b1)--node[above,fill=white,inner sep=2pt,text=red!70!black]{$0\ne1$}(b2);
\draw[red!75!black,thick] (b2)--node[above,fill=white,inner sep=2pt,text=red!70!black]{$1\ne0$}(b3);
\node[font=\scriptsize,align=center] at (2.2,-1) {Alice: $(01,11,11)$\\Discard all three records};
\end{scope}
\end{tikzpicture}
\caption{Two independent checks on a three-record path. In (a), Alice reconstructs $(01,10,11)$ and reproduces Bob's announced syndrome $(0,1)$. In (b), Alice reconstructs $(01,11,11)$: the single-bit error at vertex 2 changes both incident comparisons and triggers rejection of the entire three-record block. Edge labels compare Bob's published bit with Alice's calculated bit.}\label{fig:minimal_path}
\end{figure}

For the first example use the basis order $(X,Z)$ and let Alice prepare $(0_X,0_Z)$ in three retained rounds. Bob's choices and Alice's ideal responses are
\begin{equation}\label{eq:three_records}
 \begin{array}{c|ccc}
 i&1&2&3\\\hline
 (b_{X,i},b_{Z,i})&01&10&11\\
 (C_{X,i},C_{Z,i})&10&01&00\\
 A_i= (\widetilde b_{X,i},\widetilde b_{Z,i})&01&10&11
 \end{array}
\end{equation}
The click pairs follow $\widetilde b_\theta=a_\theta\xor(1-C_\theta)$, so in the ideal case Alice reconstructs all three records. Bob's vertex parities are $p^B=(1,1,0)$; consequently the two independent public checks on the path $1-2-3$ are
\begin{equation}\label{eq:three_syndrome}
 (s_{12},s_{23})=(0,1).
\end{equation}
If Alice reconstructs the same records, her syndrome is $(0,1)$ and the six raw coordinates pass the filter. If her second record is instead $A_2=11$ (one wrong bit relative to $10$), her parities become $(1,0,0)$ and her syndrome is $(1,0)$. Both comparisons fail and all three records are discarded. Figure~\ref{fig:minimal_path} shows the exact values; the vertex records in the drawing are not publicly disclosed.

A different alteration, $A_2=01$ instead of $10$, flips both coordinates of the second record but preserves its parity. In that case the checks still agree. Thus acceptance is a selective filter, not a proof that the two strings coincide. This undetected pattern must be included in residual reconciliation and security analysis.

\subsection{Graph representation and redundant cycle constraints}
Let $G=(V,E)$ be a fixed simple graph with $n$ vertices. We use standard graph-theoretic conventions for incidence matrices, connectivity, and spanning trees \cite{diestel}. Its binary incidence matrix $J_G\in\F^{|E|\times n}$ has two ones per edge. Bob's complete public syndrome and the discrepancy are
\begin{equation}\label{eq:graphmap}
 S=J_Gp^B,\qquad d=J_G(p^A\xor p^B)=J_Gt,
\end{equation}
where $t_i=\epsilon_{X,i}\xor\epsilon_{Z,i}$ and $\epsilon_{\theta,i}=\widetilde b_{\theta,i}\xor b_{\theta,i}$. On every edge $\{i,j\}$ the discrepancy is $d_{ij}=t_i\xor t_j$.

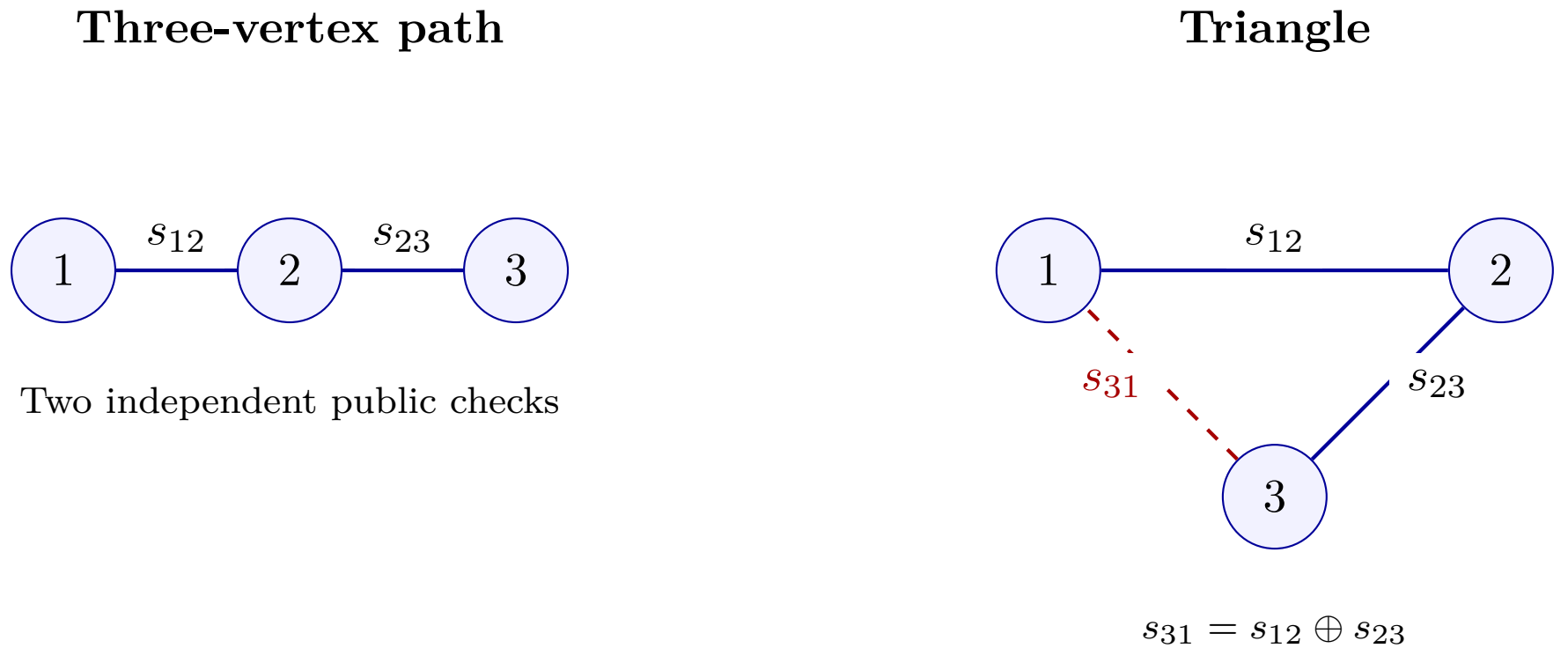
\begin{figure}[htbp]\centering
\begin{tikzpicture}[font=\small, vertex/.style={circle,draw=blue!60!black,fill=blue!5,minimum size=.78cm},edge/.style={draw=blue!60!black,thick}]
\begin{scope}
\node[font=\small\bfseries] at (1.7,1.8) {Three-vertex path};
\node[vertex] (a1) at (0,0) {$1$};\node[vertex] (a2) at (1.7,0) {$2$};\node[vertex] (a3) at (3.4,0) {$3$};
\draw[edge] (a1)--node[above,fill=white]{$s_{12}$}(a2)--node[above,fill=white]{$s_{23}$}(a3);
\node[font=\scriptsize] at (1.7,-1) {Two independent public checks};
\end{scope}
\begin{scope}[xshift=7.4cm]
\node[font=\small\bfseries] at (1.7,1.8) {Triangle};
\node[vertex] (b1) at (0,0) {$1$};\node[vertex] (b2) at (3.4,0) {$2$};\node[vertex] (b3) at (1.7,-1.7) {$3$};
\draw[edge] (b1)--node[above,fill=white]{$s_{12}$}(b2);
\draw[edge] (b2)--node[right,fill=white]{$s_{23}$}(b3);
\draw[red!65!black,thick,dashed] (b3)--node[left,fill=white,text=red!65!black]{$s_{31}$}(b1);
\node[font=\scriptsize] at (1.7,-2.7) {$s_{31}=s_{12}\xor s_{23}$};
\end{scope}
\end{tikzpicture}
\caption{The three-vertex path and triangle impose the same two independent constraints. The closing edge of the triangle is their XOR and cannot reveal any additional coordinate-error pattern. This equivalence concerns ideal syndromes, not robustness to erasures of the public checks.}\label{fig:tree_cycle}
\end{figure}

For the three vertices, a triangle adds $s_{31}=s_{12}\xor s_{23}$; in the ideal example it equals $1$. The third edge is algebraically dependent. Figure~\ref{fig:tree_cycle} compares both topologies. Any spanning tree provides exactly the same acceptance decisions as the corresponding connected graph, provided that all its edge checks are delivered and verified.

\begin{theorem}[Connected-graph acceptance kernel]\label{thm:kernel}
For a graph on $n$ vertices with $c$ connected components, $\rank(J_G)=n-c$. Define $T:\F^{2n}\to\F^n$ by $(Te)_i=e_{X,i}\xor e_{Z,i}$, and $H_G=J_GT$. Then
\begin{equation}
 \rank(H_G)=n-c,\quad
 \ker H_G=\{e:Te\text{ is constant on every component}\},\quad
 \dim\ker H_G=n+c.
\end{equation}
For a connected graph, all $n$ error parities are either zero or one. Consequently, any two connected graphs on the same vertices accept precisely the same coordinate-error patterns.
\end{theorem}
\begin{proof}
Each edge constraint requires equality of the two endpoint entries in $t$. Connectivity forces one constant on each component; hence $\dim\ker J_G=c$ and $\rank J_G=n-c$. The linear map $T$ is surjective, so $J_GT$ has the same rank. Rank--nullity and the componentwise characterization give the remaining claims.
\end{proof}

A coordinate error $(1,1)$ on one vertex has $t_i=0$ and is never detected, even by a complete graph. A single-coordinate flip changes one $t_i$ and is detected for any connected graph with $n\ge2$. For $n=1$ there is no comparison. The $n-1$ public tree checks have full independent rank; retaining a closed $n$-edge polygon spends one additional public symbol without increasing that rank.\footnote{If checks can be lost or independently corrupted on the authenticated classical channel, redundant edges may facilitate consistency monitoring; that is a distinct reliability model and not an improvement in ideal error-pattern detectability.}

\subsection{Selection of the minimum block size}
Two vertices and one edge are sufficient for nontrivial detection, so a three-vertex path is not the globally smallest possible graph. Three vertices are selected as the \emph{minimum connected size attaining the leading asymptotic coefficient one} in $Q_{\rm post}=e^2+O(e^3)$ under independent coordinate errors, as formalized in Proposition~\ref{prop:three-minimality}. This is a precise, limited design criterion, not a proof of optimal raw throughput, minimal public disclosure per useful secret bit, or maximum composable key rate. The two-vertex path is retained as an essential baseline throughout the numerical comparisons.

\section{Statistical performance of parity-graph filtering}\label{sec:laws}
\subsection{Exact acceptance and residual-error probabilities}
We distinguish acceptance, errors within accepted blocks, and errors per retained bit. $P_{\rm pass}$ is the fraction of input blocks retained. $P_{\rm und}$ is the fraction of input blocks that are both retained and erroneous. $Q_{\rm post}$ is the mean erroneous-coordinate fraction among retained blocks, while $P_{\rm block\mid pass}$ is the fraction of retained blocks containing at least one wrong coordinate. A low value of one measure does not imply that all the others are low.

Assume all $2n$ coordinate errors are independent Bernoulli variables with probability $e$. Set $q=2e(1-e)$ and $r=1-q=(1-e)^2+e^2$. On a connected graph,
\begin{align}
 P_{\mathrm{pass}}&=r^n+q^n,\label{eq:pass}\\
 P_{\mathrm{und}}&=r^n+q^n-(1-e)^{2n},\label{eq:und}\\
 Q_{\mathrm{post}}&=\frac{e^2r^{n-1}+q^n/2}{r^n+q^n},\label{eq:qpost}\\
 P_{\mathrm{block}\mid\mathrm{pass}}&=\frac{P_{\mathrm{und}}}{P_{\mathrm{pass}}}.\label{eq:block}
\end{align}
The second quantity is a joint probability over input blocks; the fourth is the error probability among retained blocks. These are different operational measures.

\begin{proof}
Acceptance requires all $t_i=0$ or all $t_i=1$, giving equation~\eqref{eq:pass}. A clean block has probability $(1-e)^{2n}$ and is always accepted; subtraction gives equation~\eqref{eq:und}. In the even branch, a vertex has two errors with conditional probability $e^2/r$. In the odd branch it has exactly one error. The expected erroneous-coordinate count divided by $2nP_{\mathrm{pass}}$ gives equation~\eqref{eq:qpost}.
\end{proof}
For fixed $n\ge3$, $P_{\mathrm{und}}=ne^2+O(e^3)$ and $Q_{\mathrm{post}}=e^2+O(e^3)$. The exceptional case $n=2$ has $P_{\mathrm{und}}=6e^2+O(e^3)$ and $Q_{\mathrm{post}}=3e^2+O(e^3)$. For $n=1$, all blocks pass and $Q_{\mathrm{post}}=e$. At $e=1/2$, $P_{\mathrm{pass}}=2^{1-n}$ and $Q_{\mathrm{post}}=1/2$; at $e=1$, all blocks pass with every coordinate wrong. Filtering alone is therefore not a correctness certificate.

\begin{proposition}[Minimal three-vertex realization of quadratic error suppression]
\label{prop:three-minimality}
Consider a connected parity graph on $n\geq 1$ two-bit records, with independent Bernoulli$(e)$ errors in every coordinate. As $e\to 0$ with $n$ fixed, the residual coordinate error probability obeys
\begin{equation}\label{eq:three-minimality}
Q_{\mathrm{post}}(n,e)=
\begin{cases}
 e, & n=1,\\
 3e^2+O(e^3), & n=2,\\
 e^2+O(e^3), & n\geq 3.
\end{cases}
\end{equation}
Hence $n=3$ is the smallest connected block attaining a unit coefficient in the leading quadratic error term. Moreover, for $0<e<1/2$, the acceptance probability $P_{\mathrm{pass}}(n,e)$ is strictly decreasing in $n$. Among the connected graphs with $n\geq3$, the three-vertex graph therefore retains the largest fraction of input blocks and requires the fewest independent parity checks, namely two. These claims do not establish optimality of the final secret-key rate.
\end{proposition}
\begin{proof}
By Eq.~\eqref{eq:qpost}, the numerator of $Q_{\mathrm{post}}$ is $e^2r^{n-1}+q^n/2$, where $r=1-2e+2e^2=1+O(e)$ and $q=2e(1-e)=2e+O(e^2)$. For fixed $n\geq2$, its first term equals $e^2+O(e^3)$, while the second is $2^{n-1}e^n+O(e^{n+1})$. The denominator $r^n+q^n$ equals $1+O(e)$. For $n=2$, the two numerator terms contribute $e^2$ and $2e^2$ at leading order, yielding $3e^2+O(e^3)$. For all fixed $n\geq3$, the second term is $O(e^3)$, leaving $e^2+O(e^3)$. For $n=1$, every block passes and the original bit-error probability is $e$.

For $0<e<1/2$, both $r$ and $q$ lie strictly between zero and one. Therefore
\begin{equation}\label{eq:three-acceptance-monotonicity}
P_{\mathrm{pass}}(n,e)-P_{\mathrm{pass}}(n+1,e)
=r^n(1-r)+q^n(1-q)>0.
\end{equation}
Finally, a connected graph on $n$ vertices has parity-check rank $n-1$ by Theorem~\ref{thm:kernel}. Thus $n=3$ is the smallest member of the unit-coefficient family, retaining the largest fraction of blocks in that family with just two independent checks.
\end{proof}

The proposition gives a precise reason to use three records as the reference design. It identifies the smallest realization of the stated asymptotic suppression property, not an optimum for all noise levels or for secret bits per emitted pulse. Larger graphs may yield different finite-$e$ residual errors, while their lower acceptance and greater public disclosure must be considered in a complete key-rate calculation.

\subsection{Correlated and nonidentical error models}
Let independent vertices have identical error-pair probabilities
\begin{equation}
 (a,b,c,d)=\Pr(00,01,10,11),\qquad r_0=a+d,\quad r_1=b+c.
\end{equation}
Within a vertex, the errors may be correlated. Directly repeating the branch calculation gives
\begin{align}
 P_{\mathrm{pass}}&=r_0^n+r_1^n,&
 P_{\mathrm{und}}&=r_0^n+r_1^n-a^n,\\
 Q_{\mathrm{post}}&=\frac{d r_0^{n-1}+r_1^n/2}{r_0^n+r_1^n}.\label{eq:corr}
\end{align}
For perfectly paired errors, $a=1-e,d=e,b=c=0$. Every block passes, $Q_{\mathrm{post}}=e$, and $P_{\mathrm{und}}=1-(1-e)^n$. Thus the quadratic suppression is a property of the independent-coordinate benchmark, not a distribution-free guarantee.

For independent but nonidentical vertices with parameters $a_i,b_i,c_i,d_i$, let $r_{0i}=a_i+d_i$ and $r_{1i}=b_i+c_i$. Then
\begin{align}
 P_{\mathrm{pass}}&=\prod_i r_{0i}+\prod_i r_{1i},&
 P_{\mathrm{und}}&=P_{\mathrm{pass}}-\prod_i a_i,\\
 Q_{\mathrm{post}}&=\frac{\sum_i d_i\prod_{j\ne i}r_{0j}+(n/2)\prod_i r_{1i}}{nP_{\mathrm{pass}}}.
\end{align}
Arbitrary dependence between vertices instead requires their joint distribution. The algebraic kernel remains valid, but product formulas do not.

\subsection{Filtering on disconnected graphs}
For component sizes $n_1,\ldots,n_c$ under independent coordinate flips,
\begin{equation}
 P_{\mathrm{pass}}=\prod_{j=1}^{c}(r^{n_j}+q^{n_j}),\qquad
 P_{\mathrm{und}}=P_{\mathrm{pass}}-(1-e)^{2n}.
\end{equation}
Given global acceptance, component independence yields
\begin{equation}
 Q_{\mathrm{post}}=\frac1n\sum_{j=1}^{c}n_j Q_{\mathrm{post}}(e,n_j).
\end{equation}
The graph rank is $n-c$. Isolated vertices have no filtering and contribute error rate $e$.

\subsection{Anchor-conditioned parity constraints}
Knowing a vertex error parity $t_a$ determines the entire vector $t$ from $J_Gt$ on a connected graph. This identifies parity errors; it does not distinguish the coordinate errors $00$ and $11$, or $01$ and $10$. Even a correctly known parity is not a certified clean record.

Three experiments must be distinguished. In the first, all $n$ vertices retain the original iid law and an extra test accepts only when $t_a=0$. The joint erroneous-acceptance probability is
\begin{equation}
 \Pr(F=1,t_a=0,\text{error})=r^n-(1-e)^{2n}.\label{eq:anchorjoint}
\end{equation}
In the second experiment, the distribution has already been conditioned on an internal anchor having $t_a=0$. The erroneous-acceptance probability in this conditioned ensemble is
\begin{equation}
 \Pr(F=1,\text{error}\mid t_a=0)
 =\frac{r^n-(1-e)^{2n}}{r}.\label{eq:anchorcond}
\end{equation}
In the third, one of the $n$ internal pairs is certified completely clean and the remaining pairs are iid. The corresponding probability is
\begin{equation}
 r^{n-1}-(1-e)^{2(n-1)}.\label{eq:anchorclean}
\end{equation}
An external clean reference attached to $n$ iid data vertices produces equation~\eqref{eq:anchorjoint} for the data, but adds a vertex and an independent constraint. Publicly revealing a parity as an anchor also consumes disclosure. Their disclosure or conditioning costs must be included. A Type-1 pair is clean in the strict loss-only model of Section~\ref{sec:physical}; its category alone is not a general certificate of correctness.

\FloatBarrier
\section{Numerical validation and retention analysis}\label{sec:results}
\subsection{Exact and simulated performance}
The independent-coordinate model admits exact comparisons for the two-node and three-node paths, and for longer connected blocks. The validation program exhaustively enumerates coordinate errors through $n=6$ and tests incidence ranks of all simple graphs through $n=5$. Its seeded Monte Carlo module samples independent errors and evaluates the actual path-edge comparisons, rather than sampling from the closed-form acceptance probability. Detector-level checks independently generate Alice's preparations, Bob's filters, and the resulting clicks. All tests use the same acceptance and conditional QBER definitions as in Section~\ref{sec:laws}. No simulation here constitutes a QKD security proof.

\begin{table}[ht]\centering\small
\caption{Exact independent-coordinate benchmark; $n=2$ and $n=3$ are the principal minimal-path candidates. $P_{\rm und}$ is the joint probability of accepting an erroneous block, whereas $Q_{\rm post}$ is the error fraction per accepted coordinate.}\begin{tabular}{@{}rrrrrr@{}}\toprule $n$ & $e$ & $P_{\rm pass}$ & $P_{\rm und}$ & $Q_{\rm post}$ & $P_{\rm block|pass}$ \\ \midrule
2 & 0.01 & 0.961184 & 0.000588 & 0.000306 & 0.000612 \\
2 & 0.05 & 0.828050 & 0.013544 & 0.008182 & 0.016356 \\
2 & 0.10 & 0.704800 & 0.048700 & 0.034620 & 0.069098 \\
3 & 0.01 & 0.941776 & 0.000296 & 0.000106 & 0.000314 \\
3 & 0.05 & 0.742075 & 0.006983 & 0.003337 & 0.009410 \\
3 & 0.10 & 0.557200 & 0.025759 & 0.017301 & 0.046229 \\
4 & 0.01 & 0.923121 & 0.000377 & 0.000102 & 0.000408 \\
4 & 0.05 & 0.670883 & 0.007463 & 0.002823 & 0.011124 \\
4 & 0.10 & 0.453172 & 0.022704 & 0.013325 & 0.050101 \\
8 & 0.01 & 0.852153 & 0.000695 & 0.000102 & 0.000816 \\
8 & 0.05 & 0.449975 & 0.009849 & 0.002762 & 0.021887 \\
8 & 0.10 & 0.204415 & 0.019113 & 0.012198 & 0.093502 \\
\bottomrule\end{tabular}
\label{tab:results}
\end{table}
\begin{figure}[ht]\centering
\includegraphics[width=.48\linewidth]{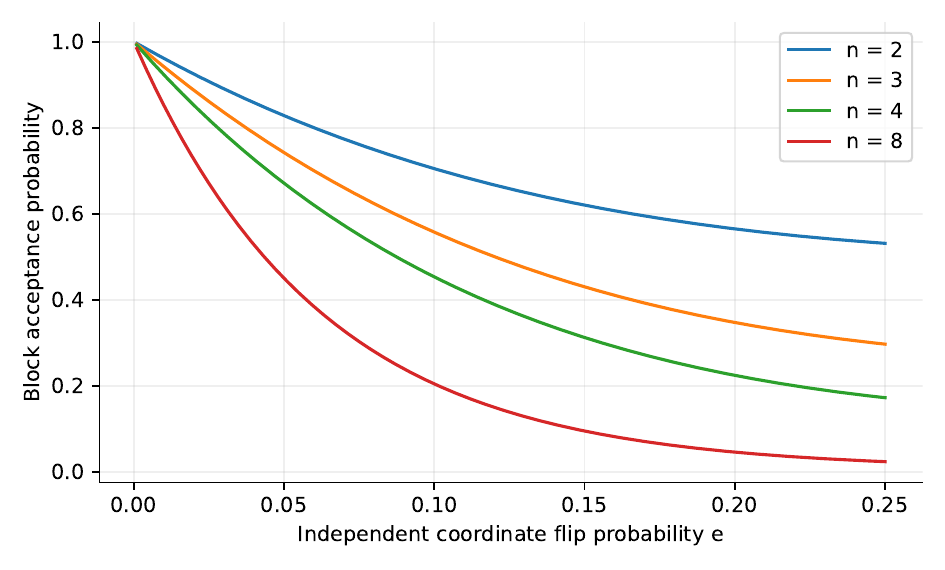}\hfill\includegraphics[width=.48\linewidth]{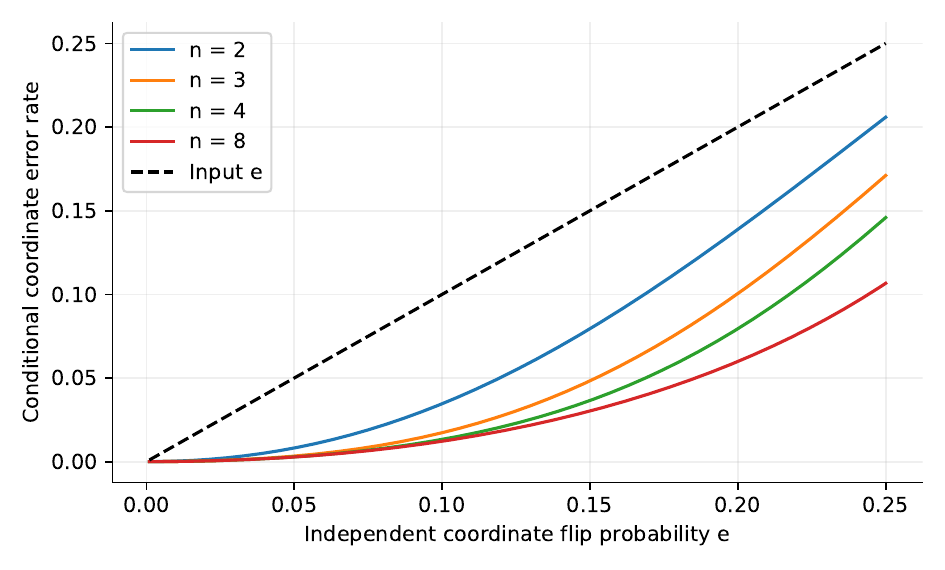}
\caption{Connected-path acceptance and residual coordinate error under independent flips. All connected graphs of the same size have the same curves. The $n=2$ path can retain substantially more blocks while having a larger leading low-error coefficient.}\label{fig:mainperformance}
\end{figure}
\begin{figure}[ht]\centering
\includegraphics[width=.48\linewidth]{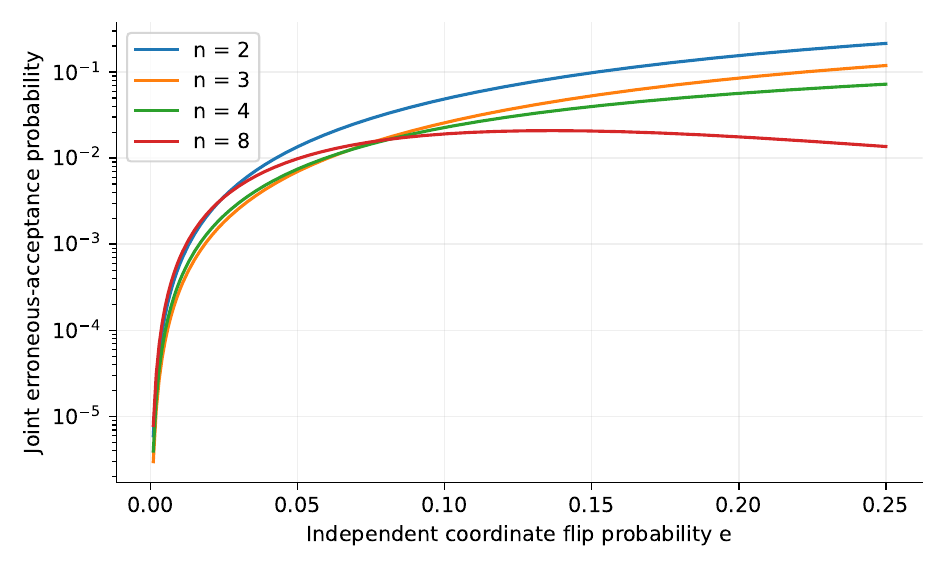}\hfill\includegraphics[width=.48\linewidth]{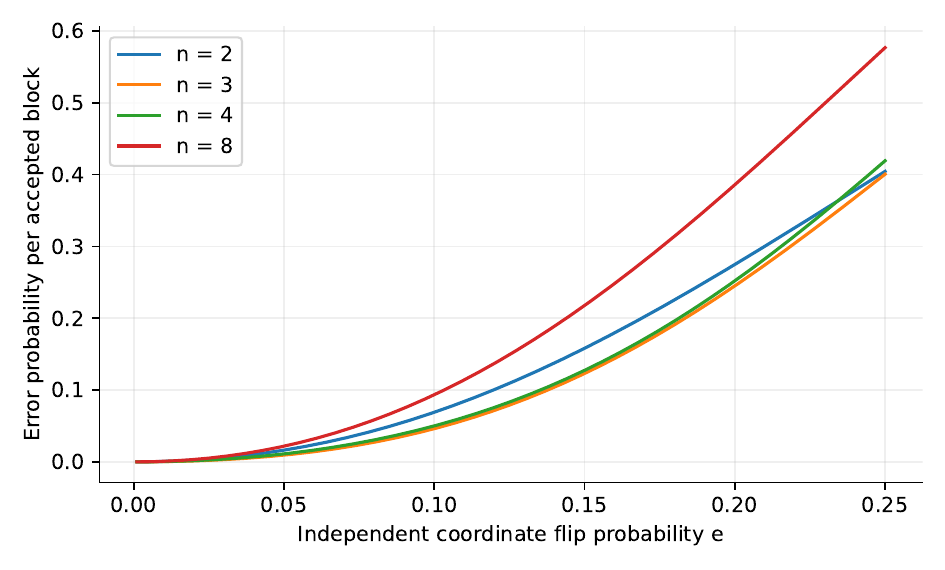}
\caption{Joint erroneous acceptance and probability of at least one erroneous coordinate conditional on passing the filter. The latter is a block-level, rather than bit-level, measure.}
\end{figure}
For $e=0.10$, a three-vertex path has $P_{\rm pass}\simeq 0.5572$ and $Q_{\rm post}\simeq 0.0173$; a two-vertex path has higher retention but larger residual QBER. As $n$ grows, $Q_{\rm post}$ approaches $e^2/r$ for $0<e<1/2$ while $P_{\rm pass}$ decreases toward zero. The larger graph cannot be identified as a more efficient reconciliation design based on QBER alone.

\begin{figure}[ht]\centering
\includegraphics[width=.48\linewidth]{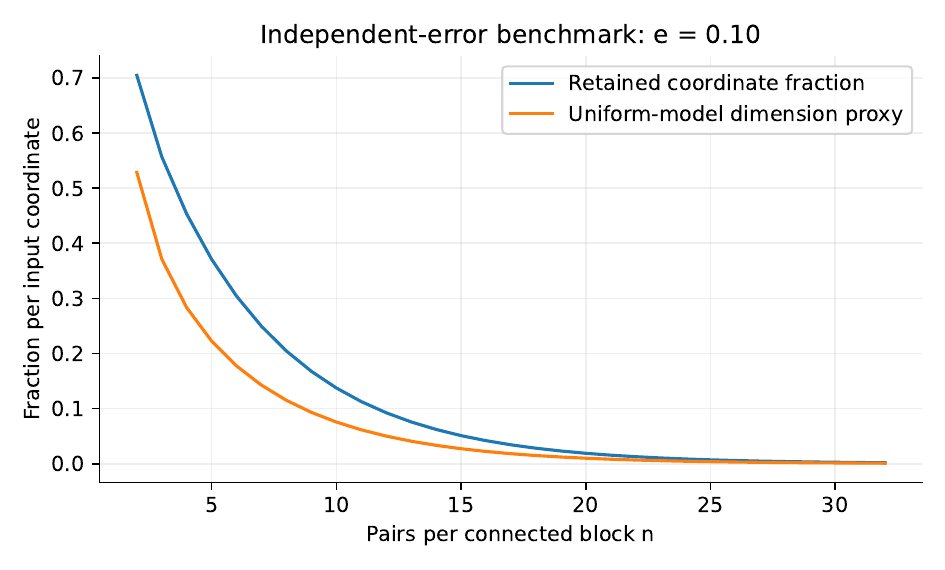}\hfill\includegraphics[width=.48\linewidth]{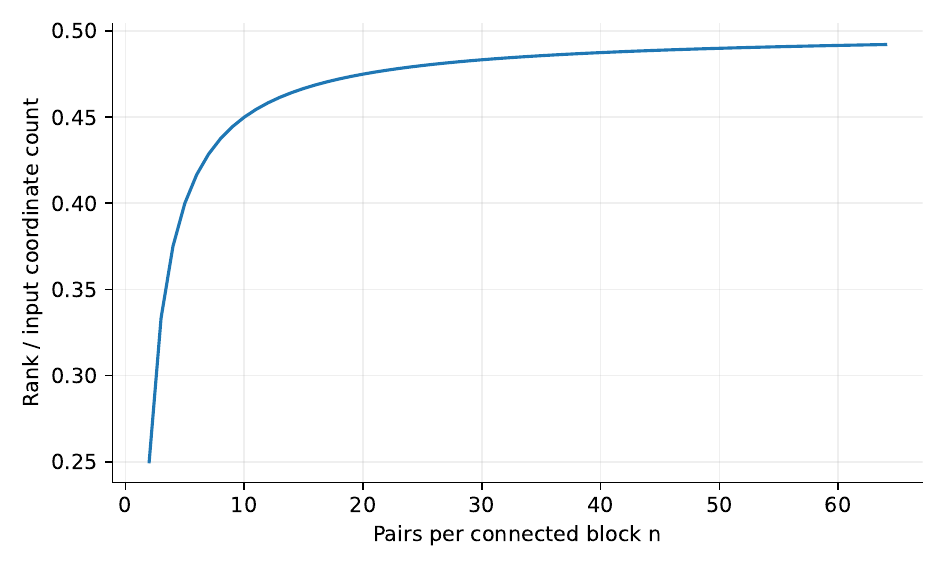}
\caption{Retention versus block size and normalized rank of the public parity checks. The displayed dimension proxy assumes uniform raw bits and does not represent a secret-key rate.}\end{figure}

\subsection{Round-trip loss and retained coordinate throughput}
Under the loss-only model of Section~\ref{sec:physical}, the independent error probability is $e=(1-\eta)/2$, where $\eta$ denotes the full unblocked round-trip detection efficiency. The table uses the three-vertex path as the reference architecture and $p_{\rm sift}=1/2$ for independently chosen basis orders.
\begin{table}[ht]\centering\small
\caption{Three-vertex path under the honest loss-only model. The final column counts retained raw-coordinate positions per emitted pulse, prior to residual reconciliation, verification, and privacy amplification.}\begin{tabular}{@{}rrrrrr@{}}\toprule $\eta$ & $e$ & $P_{\rm pass}$ & $Q_{\rm post}$ & $P_{\rm block|pass}$ & $p_{\rm sift}P_{\rm pass}$ \\ \midrule
1.0 & 0.00 & 1.000000 & 0.000000 & 0.000000 & 0.500000 \\
0.9 & 0.05 & 0.742075 & 0.003337 & 0.009410 & 0.371038 \\
0.8 & 0.10 & 0.557200 & 0.017301 & 0.046229 & 0.278600 \\
0.5 & 0.25 & 0.296875 & 0.171053 & 0.400493 & 0.148438 \\
0.2 & 0.40 & 0.251200 & 0.392357 & 0.814268 & 0.125600 \\
0.0 & 0.50 & 0.250000 & 0.500000 & 0.937500 & 0.125000 \\
\bottomrule\end{tabular}
\label{tab:physical}
\end{table}
\begin{figure}[ht]\centering
\includegraphics[width=.48\linewidth]{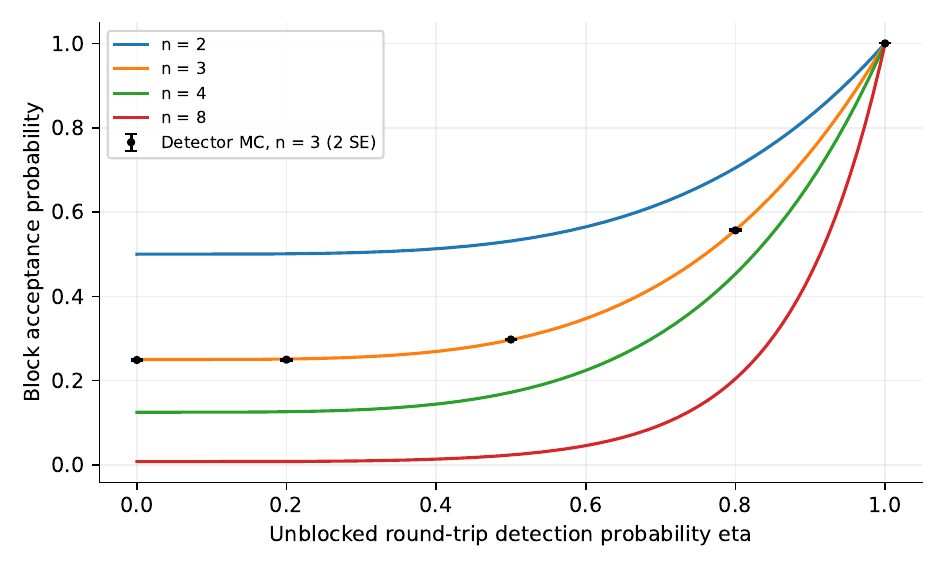}\hfill\includegraphics[width=.48\linewidth]{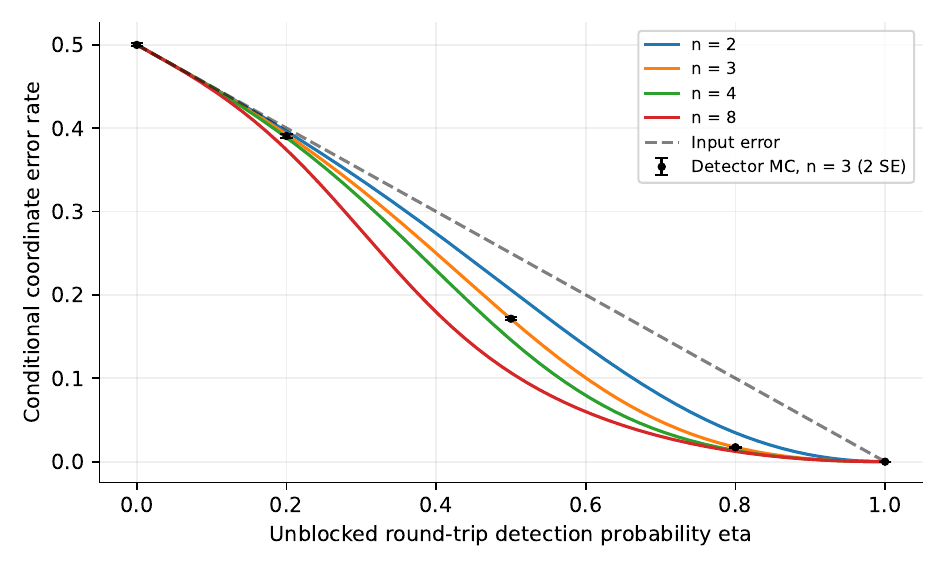}
\caption{Round-trip loss model and independent detector-level simulation for the three-vertex path. An optical efficiency of zero still permits chance passing blocks; this does not produce secret key.}\label{fig:physical}
\end{figure}
For $N$ emitted two-pulse attempts, the expected number of basis-matched records is $Np_{\rm sift}$. Grouping into blocks of size $n$ and retaining all $2n$ coordinates of an accepted block gives $2Np_{\rm sift}P_{\rm pass}$ expected raw positions before finite-block leftovers. Per emitted pulse, the raw-position fraction is $p_{\rm sift}P_{\rm pass}$. This is a bookkeeping quantity, not a rate of distilled secret bits. The two public checks per three-vertex block, acceptance announcements, later reconciliation, and privacy amplification all affect any eventual secret-key rate.

\subsection{Scope of the comparisons}
The low-error expansion distinguishes $n=2$ and $n\ge3$ but does not establish a universal ranking for finite $e$. Optimizing a practical system requires the observed error distribution, click categories, round-trip throughput, correlated errors, authenticated classical bandwidth, and an adversarial entropy bound. The longer-block and anchor-conditioned formulas remain available for that analysis; only the default topology is changed in this revision.
\FloatBarrier

\section{Public communication and security considerations}
\subsection{Linear information disclosed by the syndrome}
For a fixed graph, the image of $H_G$ has $2^{n-c}$ elements. Hence its syndrome has at most $n-c$ bits of Shannon entropy; this holds without uniformity. For a uniform Bob string $K\in\F^{2n}$ and its own syndrome $S=H_GK$,
\begin{equation}
 H(K\mid S)=n+c.\label{eq:ownentropy}
\end{equation}
This equality is not automatically true for Alice's string. For example, if Alice's uniform string $A$ and Bob's uniform string $K$ are independent, $S$ says nothing about $A$ and $H(A\mid S)=2n$.

More generally, for uniform $A$ and independent additive error $E_b$ with $K=A\xor E_b$, put $N=H_GE_b$. Since $H_GA$ is uniform on the image and independent of $N$,
\begin{equation}
 H(A\mid S)=2n-(n-c)+H(N).\label{eq:aliceentropy}
\end{equation}
Indeed $H(S)=n-c$ and $H(S\mid A)=H(N)$. On the branch $F=1$, $N=0$, and under these explicit independence assumptions the corresponding classical conditional entropy reduces to $n+c$. General quantum side information and selection-dependent inputs require a different analysis.

For fixed public metadata $M_0$, the classical register bound gives the safe min-entropy deduction
\begin{equation}
 H_{\min}^{\varepsilon}(K\mid E,M_0,S)
 \ge H_{\min}^{\varepsilon}(K\mid E,M_0)-(n-c),
\end{equation}
when $G$ is fixed by $M_0$ and the target $K$ is the same preselection string. Here $E$ is Eve's quantum system. Rank is an upper bound on the cost of this syndrome, not a measurement of Eve's total information. If the graph or pairing is chosen from private data, the public structure itself must also be included in $M_0$.

\subsection{Restricted optical-access security model}\label{sec:access}
The intended Loop-Back deployment consists of a fixed station operated by Alice, combining optical transmission and reception, and a portable terminal operated by Bob. Alice controls source preparation, detector calibration, pulse ordering, and the acceptance windows for the returning signals. Bob's trusted terminal selects two polarization-transmission branches per record without recording a quantum measurement. Public basis-order information and classical reconciliation are authenticated. The allowed temporal and optical modes are specified by the station; an accepted click must be consistent with the expected return window. Absence of a click remains a legitimate outcome and provides no direct time-of-arrival evidence.

For the minimal security statement, the implementation-independent description of Eve's access is a single bound on her \emph{pre-filter} guessing probability,
\begin{equation}\label{eq:gmodel}
P_{\rm guess}(K\mid E,M_0)\le g,\qquad 2^{-6}\le g\le 1,
\end{equation}
where $K$ is the six-bit filter-selection block and $M_0$ includes every public announcement preceding graph disclosure. The spatial and temporal restrictions define the intended operation but do not alone prove $g<1$. A secure optical model must supply the bound. In the particular trusted iid passive-tap channel analyzed below, $g=((1+\lambda)/2)^6$; more general optical interactions require a separately justified bound and may yield $g=1$.

This form keeps the argument reusable: a different optical channel changes the bound $g$, but not the graph theorem or the privacy-amplification step. The station's locality and timing constrain some delay-inducing relays, but do not exclude on-path zero-delay interactions or active probing in a legitimate input mode. The classical checks are applied only after the entire forward--return optical exchange.

\subsection{Conditional security against quantum side information}\label{sec:security}
Let $K\in\{0,1\}^6$ denote Bob's three paired records and $A$ Alice's inferred string. Let $E$ be an arbitrary quantum system held by Eve after the entire optical exchange and let $M_0$ include the authenticated public transcript preceding graph disclosure, including the basis orders and any public selection decisions. No independence or bitwise noise assumptions are made for this theorem. For the path $1-2-3$, Bob releases the two-bit syndrome $S=HK$, with
\begin{equation}\label{eq:Hsecurity}
H=\begin{pmatrix}1&1&1&1&0&0\\0&0&1&1&1&1\end{pmatrix}.
\end{equation}
Alice announces $F=1$ if and only if $HA=S$; $p_{\mathrm{acc}}=\Pr(F=1)$ is the probability of this publicly declared event.

\begin{theorem}[Accepted-branch min-entropy bound]\label{thm:conditional}
For an arbitrary classical--quantum state $\rho_{KAE M_0}$, if $p_{\mathrm{acc}}>0$ and
\begin{equation}\label{eq:preent}
H_{\min}(K\mid E M_0)_\rho\ge h,
\end{equation}
then the normalized accepted state satisfies
\begin{equation}\label{eq:conditionalmin}
H_{\min}(K\mid E M_0 S,F=1)\ge h-2+\log_2p_{\mathrm{acc}}.
\end{equation}
The bound holds even when Eve's quantum side information is correlated with Alice's errors or the public acceptance event.
\end{theorem}
Equivalently, inserting the certified bound $g$ from Eq.~\eqref{eq:gmodel} gives
\begin{equation}\label{eq:gbound}
\begin{aligned}
P_{\rm guess}(K\mid E,M_0,S,F=1)&\le\min\{1,4g/p_{\rm acc}\},\\
H_{\min}(K\mid E,M_0,S,F=1)&\ge\max\{0,-\log_2(4g/p_{\rm acc})\}.
\end{aligned}
\end{equation}
The sufficient condition for a strictly positive accepted-block lower bound is $4g<p_{\rm acc}$; positive extractable key length further requires a joint many-block bound and the accounting of reconciliation, verification, authentication, and finite-size terms. A proof based on quantum guessing probabilities is given in Section~\ref{app:security} of Appendix~\ref{app:proofs}. Equation~\eqref{eq:conditionalmin} is a sufficient conditional security reduction, not a lower bound on the pre-filter entropy $h$ inferred from an observed QBER. Its right-hand side can be negative and hence uninformative.

After acceptance, let $C_{\mathrm{EC}}$ and $C_{\mathrm{ver}}$ denote authenticated reconciliation and verification transcripts and let $\lambda$ upper-bound their total log-cardinality (or another proved conditional entropy leakage bound). If Alice and Bob reconcile a sufficiently long accepted string and subsequently apply a randomly seeded two-universal hash, a key of length $\ell$ can be extracted with secrecy error at most $\varepsilon_{\mathrm{PA}}$ under the unsmoothed convention provided
\begin{equation}\label{eq:keylength}
\ell\le H_{\min}(K_{\mathrm{acc}}\mid E,M_0,S,F=1)-\lambda
-2\log_2\!\left(\frac{1}{2\varepsilon_{\mathrm{PA}}}\right).
\end{equation}
Here $K_{\mathrm{acc}}$ denotes the actual concatenation of selected blocks; a valid joint min-entropy estimate is needed for that concatenation. Per-block entropy bounds may not be added without an independence, entropy-accumulation, or equivalent argument. Standard correctness and authentication failures contribute separately to the complete failure budget \cite{tom,dw,leftover}.

\begin{proposition}[Insufficiency of the optical abstraction]\label{prop:nogo}
The ideal preparation/filter/return rule, spatial colocation of Alice's transmitter and receiver, timing-window constraints, and any fixed honest-operation QBER and graph-acceptance probabilities alone do not imply $H_{\min}(K\mid E,M_0)>0$. In particular, these observables admit an extension in which Eve possesses a perfectly distinguishable record of Bob's filter settings while the Alice--Bob statistics are unchanged.
\end{proposition}
The formal extension and its physical interpretation are developed in Section~\ref{app:security} of Appendix~\ref{app:proofs}. No positive unconditional secret-key rate for the specified Loop-Back device follows from Theorem~\ref{thm:conditional} alone. Active incoming-light probing of a modulating/reflective Bob station is a relevant mechanism to rule out or bound \cite{trojan,gisin2006}.

\subsection{Explicit restricted-channel optical bound}\label{sec:passive}
A positive preselection entropy can be derived once the optical adversary is restricted sufficiently to specify Eve's returned states. We analyze a \emph{trusted passive-tap model}: the source emits one photon per slot, no adversarial light is injected, Bob has ideal basis-matched single-output filters, and a lossless beam splitter transfers the transmitted photon to Eve with probability $\lambda$ and otherwise to Alice. The six slot interactions are independent. Eve retains all tapped optical modes in quantum memory. Conservatively, the entire six-bit Alice preparation string $a$ is given to Eve. These are assumptions about the permitted attack, not consequences of observed attenuation.

For a slot with $d=b\xor a$, Eve's conditional states in the vacuum--single-photon sector are
\begin{equation}\label{eq:evetapstates}
 \rho_E^{d=0}=(1-\lambda)|\mathrm{vac}\rangle\langle\mathrm{vac}|+\lambda|1\rangle\langle1|,\qquad
 \rho_E^{d=1}=|\mathrm{vac}\rangle\langle\mathrm{vac}|.
\end{equation}
Their optimal distinguishing probability for equiprobable $d$ is $(1+\lambda)/2$; for six independent bits,
\begin{equation}\label{eq:preoptical}
 h_6=H_{\min}(K\mid E,a)=6\left[1-\log_2(1+\lambda)\right].
\end{equation}
In this channel the tap causes an error only on a matched-value ($d=0$) pulse, so the average single-bit inference error is $e=\lambda/2$. With $q=2e(1-e)$ and $r=1-q$, the three-node acceptance probability is $p_{\rm acc}=r^3+q^3$. The general postselection theorem yields
\begin{equation}\label{eq:passivebound}
 H_{\min}(K\mid E,a,S,F=1)\ge 6[1-\log_2(1+\lambda)]-2+\log_2(r^3+q^3).
\end{equation}
This bound accounts for correlation of Eve's tap outcomes with the acceptance event: no independence assumption between Eve and $F$ has been introduced. For $\lambda=0.10$, $p_{\rm acc}=0.742075$ and the right-hand side is $2.744616$ bits per accepted six-bit block.

The restricted channel permits a tighter exact calculation. Write $t_i=1$ when Eve captures the photon in slot $i$ and otherwise $t_i=0$. Given $a$, use coordinates $d=K\xor a$, and denote by $H$ the two-check matrix in Eq.~\eqref{eq:Hsecurity}. In this model Alice's reconstruction error is \emph{exactly} $t$, and hence $F=1$ if and only if $Ht=0$; Eve learns $t$ from her orthogonal vacuum--photon record. The whole-block guessing probability after disclosure is
\begin{equation}\label{eq:exactguess}
 P_{\rm guess}(K\mid E,a,S,F=1)=\frac1{64p_{\rm acc}}
 \sum_{\substack{t\in\{0,1\}^6\\Ht=0}}\ \sum_{s\in\{0,1\}^2}
 \max_{\substack{d\in\{0,1\}^6\\Hd=s,\;t\wedge d=0}}
 \lambda^{|t|}(1-\lambda)^{6-|t|-|d|}.
\end{equation}
Here $|d|$ is Hamming weight, $t\wedge d=0$ means coordinatewise disjoint support, and $s$ is the syndrome after subtracting the known shift $Ha$. Empty maxima are defined as zero. At $\lambda=0.10$ Eq.~\eqref{eq:exactguess} gives $P_{\rm guess}=0.08082217$, or
\begin{equation}
 H_{\min}(K\mid E,a,S,F=1)=3.629105\ \text{bits},
\end{equation}
which is strictly stronger than Eq.~\eqref{eq:passivebound}. The proof and exhaustive enumeration are given in Section~\ref{app:passiveproof} of Appendix~\ref{app:proofs}.

For the same physically specified tap channel, the honest $K$--$A$ error \emph{is correlated with Bob's bits when $a$ is fixed}. Therefore Eq.~\eqref{eq:hexact}, which assumes additive noise independent of $K$, must not be substituted as an equality for this model. The exact classical reconciliation entropy follows from the finite distributions in Section~\ref{app:passiveproof} of Appendix~\ref{app:proofs}; at $\lambda=0.10$ it is $H(K\mid A,a,S,F=1)=0.080208$ bits per accepted block. A conservative bound remains $H(K\mid A,a,S,F=1)\le H(t\mid F=1)=0.098691$ bits. For independent accepted blocks with asymptotically ideal reconciliation, the restricted-channel one-way entropy difference gives
\begin{equation}\label{eq:passiverate}
 R_{3}^{\mathrm{tap}}=\frac{p_{\rm sift}p_{\rm acc}}{3}
 [H(K\mid E,a,S,F=1)-H(K\mid A,a,S,F=1)]_+,
\end{equation}
measured in secret bits per emitted two-pulse pair. This is an asymptotic \emph{restricted-channel} formula, not a composable security statement for unrestricted optical attacks. With $p_{\rm sift}=1/2$, $\lambda=0.10$, and the exact entropies above, Eq.~\eqref{eq:passiverate} yields $0.480845$ secret bits per emitted pair. A conservative min-entropy/error-entropy calculation yields $0.327246$ bits per emitted pair as an asymptotic benchmark, before finite-size and authentication costs. These figures are not experimentally certified rates.

\begin{table}[htbp]\centering\small
\caption{Exact accepted-branch quantities for the trusted iid passive-tap model, with the six preparation bits granted to Eve. The key-rate column includes $p_{\rm sift}=1/2$ and is expressed per emitted two-pulse pair; it is not a rate certified against general attacks.}
\label{tab:tap}
\begin{tabular}{@{}rrrrrr@{}}\toprule
$\lambda$ & $p_{\rm acc}$ & lower $H_{\min}$ & exact $H_{\min}$ & $H(K\mid A,a,S,F=1)$ & $R_3^{\rm tap}$\\\midrule
0.00&1.0000&4.0000&4.0000&0.0000&0.6667\\
0.05&0.8609&3.3615&3.8269&0.0218&0.5698\\
0.10&0.7421&2.7446&3.6291&0.0802&0.4808\\
0.20&0.5572&1.5781&3.1508&0.3106&0.3281\\
\bottomrule\end{tabular}
\end{table}

\subsection{Optical assumptions and certification boundary}
A measured optical loss is not an experimental certificate of Eve's tap parameter $\lambda$: active replacement, additional incoming modes, injected multiphoton probes, wavelength-dependent reflections, and forward--return correlations can violate the passive-tap channel while preserving the same Alice--Bob click frequencies. An optical implementation requires a characterization of all modes entering and leaving Bob and a security reduction that bounds the distinguishability of Bob's setting-dependent returned states \emph{for every allowed attack}. Studies of Trojan-horse attacks on QKD systems illustrate the importance of controlling injected light and unintended return modes \cite{gisin2006,trojan}. Monitor detectors, optical isolation, spectral filtering, and out-of-band tests are possible engineering controls, but their effectiveness must be quantified for the particular device. Such controls motivate a restricted model but do not, without quantitative bounds and an attack-uniform proof, establish its validity. Consequently the positive formulas in this subsection close security only for the declared optical tap family. For unrestricted attacks the formal insufficiency result of Proposition~\ref{prop:nogo} remains in force.

\subsection{Exact reconciliation entropy under independent errors}
For the \emph{honest}, independent symmetric-error model, uniform $K$ and independent error vector $N=A\xor K$, the pass event depends only on $N$. Writing $q=2e(1-e)$, $r=1-q$, $P=r^n+q^n$, and $w_1=q^n/P$, the uncertainty of Bob given Alice on a passed block is exactly
\begin{equation}\label{eq:hexact}
H(K\mid A,S,F=1)=H(N\mid F=1)
=h_2(w_1)+n(1-w_1)h_2(e^2/r)+nw_1.
\end{equation}
For $e=0.1$ and $n=3$ this equals $0.397339$ bits per accepted six-bit block, with $p_{\mathrm{acc}}=0.5572$. This is an estimate of the theoretical reconciliation cost in a declared benign stochastic model; it does not bound Eve's information. A proof and exhaustive six-bit validation are included in Section~\ref{app:security} of Appendix~\ref{app:proofs}.

\subsection{Residual reconciliation and key verification}
After reconstruction and basis sifting, the remaining operations are performed in the following order:
\begin{enumerate}
\item Fix the graph independently of unrevealed key data, or account for the information disclosed by its selection.
\item Publish Bob's syndrome, compare edge values, and communicate the retained block indices.
\item Estimate the relevant parameters with the selection rule included, then reconcile residual discrepancies.
\item Verify equality with a fresh two-universal hash and apply privacy amplification using a proved entropy bound.
\end{enumerate} A $\tau$-bit verification tag bounds collision for distinct fixed strings by $2^{-\tau}$ when its seed is independent. Its tag and all residual syndromes must be counted. Finally, privacy amplification uses an independently chosen seed and a proved entropy bound. Authentication failure is another explicit part of the total failure budget.

A bare SHA-256 equality check is not the same as an information-theoretic two-universal verification argument. Nor does accepting a parity block ensure equal strings. In particular, a single within-pair double flip passes every graph check. Keeping accepted blocks without residual correction or final rejection does not establish correctness.

\subsection{Relation to repetition-code advantage distillation}
The path syndrome is the usual parity-check representation of a length-$n$ binary repetition code applied to the \emph{vertex parity sequence} $p=(p_1,\ldots,p_n)$. Its pass condition is that the error-parity sequence is constant. The broader secret-key-agreement perspective and modern QKD applications of repetition-code advantage distillation have been studied previously \cite{maurer1993,tan,adpractical}; the present mathematical postselection should not be claimed as a new coding principle. Its physically specific input consists of two-coordinate records inferred from selective-polarizer, round-trip, paired-state measurements.

An important difference from a conventional repetition-code distillation step is the output: the present protocol retains both coordinates of every accepted vertex, whereas familiar repetition-code variants commonly retain one new bit per accepted block. The extra per-vertex even-parity error $(1,1)$ survives every graph test and has to be addressed by subsequent processing. Therefore equal acceptance rules on $p$ do not imply the same output entropy, final secret-key rate, or security proof. Rank $n-1$ describes the syndrome disclosure for a fixed connected graph; it is not a lower bound on secrecy against Eve.

Cascade, Winnow and LDPC-based reconciliation instead use interactive searches, Hamming syndromes, or sparse-code syndrome decoding to remove residual discrepancies \cite{cascade,winnow,ldpc}. The path rule performs error detection and block postselection; it neither locates a uniquely correct error pattern nor decodes retained strings. Comparisons with other Loop-Back variants must be made with final secret bits per emitted optical pulse at matched optical and adversarial assumptions, not only with postselected error rates.

Repeatedly combining the same raw events cannot create new independent entropy. In any hybrid construction, all reused records, disclosed parity equations, and public selections must be accounted for jointly.

\FloatBarrier

\section{Operational interpretation and research directions}\label{sec:outlook}
The practical objective of Loop-Back is a fixed optical station that prepares and detects quantum signals while portable terminals implement lightweight optical operations. Alice's source, receiver, timing electronics, and authenticated classical interface may be reused across successive user sessions. This architecture motivates a community-facing quantum service rather than independent full QKD transceivers at every user device \cite{loop,recent}. The present work analyzes one station--one portable-user link; it does not demonstrate simultaneous multiuser capacity, scheduling performance, or secret-key delivery between two users.

\subsection{Integration with error reconciliation}
The parity graph is a detection and selection procedure, not a standalone error-correcting code. When a block fails any edge check, the entire block is discarded; when it passes, undetected double-coordinate errors can remain. Therefore correctness still requires reconciliation and verification. Winnow uses Hamming-code syndromes to correct some mismatches, whereas LDPC-based schemes use sparse parity constraints and a decoder \cite{winnow,ldpc}. Such procedures may operate directly on the sifted paired records or after graph filtering. A hybrid construction must charge both its initial parity announcements and the later reconciliation leakage, and must measure the fraction of records discarded during filtering.

The pertinent comparison is between two complete processing chains under the \emph{same} optical observations and adversarial model:
\begin{equation}\label{eq:future_comparison}
\begin{aligned}
&\text{paired-state acquisition}\;\longrightarrow\;\text{reconciliation}\;\longrightarrow\;\text{privacy amplification},\\
&\text{paired-state acquisition}\;\longrightarrow\;\text{graph filtering}\;\longrightarrow\;\text{reconciliation}\;\longrightarrow\;\text{privacy amplification}.
\end{aligned}
\end{equation}
The performance criterion is \emph{composable secret bits per emitted optical pulse}, including basis sifting, acceptance probability, disclosed syndromes, reconciliation, authentication, and finite-key penalties. The error-rate improvement established here does not imply that the second chain yields more secret bits. Comparing path lengths and correction codes on this common basis is a concrete research task rather than a performance result of the present manuscript.

\subsection{Extension to two portable users}
A second direction replaces the one-user exchange by an ordered optical route through two portable terminals, schematically $A\rightarrow B_1\rightarrow B_2\rightarrow A$. The station would still supply quantum states, timing and detection, while $B_1$ and $B_2$ would aim to establish a secret key with each other using local operations. Multi-node Loop-Back architectures provide motivation for this topology \cite{loop,recent}; separate experiments on multiparty quantum secret sharing and conference key agreement illustrate a related network-level goal \cite{mpqss}, but the selective-polarizer channel and two-bit-record statistics derived here cannot be transferred to sequential portable devices without a new optical model. In particular, multiple filters in series may produce different transmission and vacuum-event probabilities from the single-user model; a unitary rotational realization is a distinct option.

The cryptographic objective also changes: an honest-but-curious station Alice must not learn the users' eventual key. If $K_{12}$ denotes their provisional common string and $E$ an external adversary, the requisite pre-filter input to the general theorem would be a bound such as
\begin{equation}\label{eq:multiuser_entropy_target}
P_{\rm guess}(K_{12}\mid E,A_{\rm station},M_0)\leq g_{12}<1,
\end{equation}
where $A_{\rm station}$ includes Alice's preparation choices, measurements and other station records. The conditional-security theorem of Section~\ref{sec:security} can then account for a public syndrome of rank $r$ and the associated acceptance event without changing its proof. Deriving a nontrivial $g_{12}$, establishing correctness between $B_1$ and $B_2$, and addressing a malicious rather than merely curious station are separate open problems. In particular, uncertainty about the users' individual operations does not by itself establish secrecy of a shared function of those operations.

Both extensions serve a longer-term objective: a shared quantum station supporting a community of lightweight mobile terminals and, eventually, user-to-user key establishment. They are proposed research directions, not capabilities or security guarantees established by the present single-user protocol.

\section{Conclusions}
The paired-state Loop-Back exchange provides two-bit records using a fixed optical station and a portable user terminal. Bob's filter selections are inferred from return-window detections, including non-click outcomes in the ideal channel. This arrangement retains the central practical motivation of Loop-Back: costly sources, detectors, and timing electronics remain at Alice's station, while the user's optical operations are simple.

We have shown that a connected parity graph needs only a spanning tree of checks. For three records, the path $1-2-3$ has two independent public parities and exactly the same acceptance set as the closed triangle. This is not merely a graphical simplification: it identifies the minimum independent information needed by the test. The resulting exact laws give $Q_{\rm post}=e^2+O(e^3)$ for three or more records and $3e^2+O(e^3)$ for two records under independent bit errors. Three nodes are therefore the smallest block attaining the first leading coefficient. We have also characterized the remaining error patterns and the fraction of blocks sacrificed to obtain this improvement.

The security analysis makes the public cost of the test explicit. The syndrome and acceptance announcement obey a conditional min-entropy bound against arbitrary quantum side information, provided that the optical exchange supplies a certified pre-filter guessing bound. A fully specified passive-tap channel yields both a positive bound and an exact calculation of Eve's accepted-block knowledge. These results establish security for that defined channel and authenticated postprocessing; they do not extend automatically to unrestricted active probing of Bob's device. The distinction identifies the next physical problem to solve without changing the graph theorem.

The graph-based advantage-distillation step is intended as a selection stage before residual error reconciliation, not as an alternative decoder to Winnow or LDPC. Its net value will be determined by comparing full secret-key rates with and without filtering under the same channel and adversarial assumptions. A second extension will allow two portable users to access Alice's shared optical station and establish a key confidential from the station itself. The single-user results provide the graph and postselection framework for that study, while the two-user optical correlations and secrecy bound remain to be derived. Together, these directions advance the larger aim of a community-accessible Loop-Back quantum infrastructure.

\appendix
\section{Parity filtering: algebraic and statistical details}\label{app:algebra}
\paragraph{Three-record error patterns.}
In the three-record example of Section~\ref{sec:polygon}, Bob holds $(01,10,11)$ and has syndrome $(0,1)$. If Alice holds $(01,01,11)$, the middle record is the coordinate complement of Bob's $10$; its parity remains one, so the syndrome remains $(0,1)$ and the erroneous block passes. If Alice instead holds $(01,11,11)$, one coordinate differs at the middle vertex and Alice's syndrome is $(1,0)$, so the block fails. An error of odd coordinate parity at every vertex also passes, since $t=(1,1,1)$ belongs to the connected-graph kernel. There are $2^{n+1}=16$ accepted coordinate-error vectors for $n=3$ out of $2^{2n}=64$ possible vectors; their probabilities are not equal unless $e=1/2$.

\paragraph{Syndrome disclosure and conditioning.}
For a connected graph and uniform independent $A,K\in\F^{2n}$, Bob's syndrome has entropy $n-1$, but $I(A;S)=0$. Thus $H(A\mid S)=2n$, not $n+1$. Taking iid additive noise with $e=1/2$ realizes this counterexample. This does not contradict equation~\eqref{eq:ownentropy}, which concerns $K$ and its own syndrome, or the accepted-branch calculation under its explicit assumptions.

\paragraph{Accepted independent-error blocks.}
\begin{proposition}\label{prop:honest}
If $K$ is uniform and independent of the iid Bernoulli$(e)$ error vector $N=A\xor K$, then
\begin{equation}
H(K\mid A,S,F=1)=h_2(w_1)+n(1-w_1)h_2(e^2/r)+n w_1,
\end{equation}
where $q=2e(1-e)$, $r=1-q$, $P=r^n+q^n$, $w_1=q^n/P$, and $h_2(x)=-x\log_2x-(1-x)\log_2(1-x)$.
\end{proposition}
\begin{proof}
For each vertex, the two coordinate errors have even parity with probability $r=(1-e)^2+e^2$ and odd parity with probability $q=2e(1-e)$. Since the graph is connected, acceptance requires that every vertex have even error parity, or that every vertex have odd error parity. The respective probabilities are $r^n$ and $q^n$, hence $P=r^n+q^n$. Given acceptance, the odd branch occurs with probability $w_1=q^n/P$. On the all-even branch each error pair equals $(0,0)$ or $(1,1)$ with conditional probabilities $(1-e)^2/r$ and $e^2/r$, giving $n h_2(e^2/r)$ bits. On the all-odd branch each pair is $(1,0)$ or $(0,1)$ with equal probability, giving $n$ bits. The entropy of the disjoint mixture adds $h_2(w_1)$ and the weighted branch entropies. Uniform independent $K$ makes $A=K\xor N$ independent of $N$; since $S=HK=HA$ on $F=1$, knowledge of $S$ adds no information to $(A,F=1)$. Finally, $K=A\xor N$, so $H(K\mid A,S,F=1)=H(N\mid F=1)$, which proves the formula.
\end{proof}

This quantity is a reconciliation entropy in an honest model, not an adversarial entropy bound. For $n=3$ and $e=0.10$, direct enumeration over all $2^6$ error vectors gives $P=0.5572$ and $H(N\mid F=1)=0.3973389579$ bits.

\section{Security proofs}\label{app:proofs}
\subsection{General postselection and extraction}\label{app:security}
\paragraph{Model and min-entropy conventions.}
All logarithms are base two. The input to postprocessing is the normalized state
\begin{equation}\label{eq:cqsource}
\rho_{KAE M_0}=\sum_{k,a,m}p(k,a,m)\,|k,a,m\rangle\!\langle k,a,m|\otimes\rho_E^{k,a,m}.
\end{equation}
The variable $K$ is the six-bit raw filter-selection string, $A$ is Alice's six-bit estimate, $M_0$ is the public classical transcript prior to parity disclosure, and $E$ is an unrestricted quantum system. The order of classical tensor factors is immaterial. We impose no conditional independence of $E$ and $A$. The (non-smooth) conditional min-entropy of a classical variable is
\begin{equation}
H_{\min}(K\mid E M_0):=-\log_2P_{\rm guess}(K\mid E M_0),
\end{equation}
where $P_{\rm guess}$ maximizes the average success probability of a quantum measurement on Eve's side information, including the known classical $M_0$.

The graph check is a deterministic classical function of $(K,A)$:
\begin{equation}
S=HK,\qquad F={\bf1}_{\{HA=HK\}}.
\end{equation}
In the proof below Eve is given both public variables, so their possible correlations with her quantum system are fully covered. Let $p=\Pr(F=1)>0$.

\paragraph{Syndrome disclosure lemma.}
\begin{proposition}\label{prop:syndrome}
Let $X$ be classical, $Q$ quantum or classical--quantum, and $S=f(X)$ a public variable taking at most $D$ values. Then
\begin{equation}\label{eq:syndromebound}
P_{\rm guess}(X\mid Q,S)\le D\,P_{\rm guess}(X\mid Q),
\end{equation}
and $H_{\min}(X\mid Q,S)\ge H_{\min}(X\mid Q)-\log_2D$.
\end{proposition}
\begin{proof}
For each syndrome value $s$, let $\{M_x^s\}_x$ be a measurement optimal for guessing $X$ when $S=s$ is known. Construct a valid unconditional guessing measurement by first choosing $s$ uniformly from the $D$ advertised values and then applying $\{M_x^s\}_x$. When the randomly chosen $s$ equals the actual $f(X)$, which occurs with probability $1/D$, this strategy achieves $1/D$ times the conditional guessing success probability (the remaining contributions are nonnegative). Since the optimal unconditional measurement can only do better, inequality~\eqref{eq:syndromebound} follows. The same argument works with a classical public transcript incorporated in $Q$, conditioning on its value. Taking negative logarithms gives the entropy statement.
\end{proof}

For the three-node path, $S\in\F^2$ and hence $D\le4$. Its matrix has rank two, so all four values are possible for uniform input. The bound by two is valid without uniformity and regardless of correlation with Eve.

\paragraph{Acceptance conditioning lemma.}
\begin{proposition}\label{prop:accept}
For a cq-state with a publicly revealed binary event $F$ and $p=\Pr(F=1)>0$,
\begin{equation}\label{eq:acceptbound}
P_{\rm guess}(X\mid Q,F=1)\le \frac{P_{\rm guess}(X\mid Q)}{p}.
\end{equation}
The system $Q$ may itself include previously announced classical transcripts.
\end{proposition}
\begin{proof}
Write the unnormalized accepted cq operator as $\widetilde\rho_{XQ}^{\rm acc}=p\rho_{XQ\mid F=1}$. Since the original cq operator is the sum of the accepted and rejected positive operators, $0\le\widetilde\rho_{XQ}^{\rm acc}\le\rho_{XQ}$ in the block-diagonal operator order. Every guessing measurement has a positive success operator. Its success probability on the accepted substate is therefore no larger than on the original state. Maximizing over all measurements for the accepted state and dividing by $p$ proves~\eqref{eq:acceptbound}. The argument does not require $F$ to be independent of $X$ or $Q$.
\end{proof}

\paragraph{Proof of the accepted-branch theorem.}
\begin{proof}[Proof of Theorem~\ref{thm:conditional}]
Apply Proposition~\ref{prop:syndrome} with $X=K$, $Q=(E,M_0)$ and $D=4$:
\begin{equation}
P_{\rm guess}(K\mid E,M_0,S)\le 4\,2^{-h}.
\end{equation}
Then apply Proposition~\ref{prop:accept} to the accepted event, with $Q=(E,M_0,S)$:
\begin{equation}
P_{\rm guess}(K\mid E,M_0,S,F=1)\le\frac{4}{p_{\rm acc}}2^{-h}.
\end{equation}
Taking negative logarithms establishes
\begin{equation}
H_{\min}(K\mid E,M_0,S,F=1)\ge h-2+\log_2p_{\rm acc}.
\end{equation}
No step imposes an iid assumption, a classical restriction on Eve, or independence between accepted blocks. This concludes the claimed one-block, non-smooth statement.
\end{proof}

\begin{remark}[Role of the acceptance probability]
The numerical value $p_{\rm acc}=0.5572$ derived from independent coordinate errors at $e=0.1$ is a property of the honest stochastic model. In an adversarial implementation, $p_{\rm acc}$ must be bounded from the actual joint state or conservatively estimated from authenticated public data; it cannot be inserted into a security bound as an invariant of Eve's attack. The term $\log_2 p_{\rm acc}$ may be negative and reflects the conditioning penalty, not a gain in secrecy.
\end{remark}

\paragraph{Privacy amplification and correctness.}
\begin{proposition}[Conditional composable extraction]\label{prop:extract}
For the actual accepted concatenated string $K_{\rm acc}$, suppose all public announcements preceding reconciliation are included in $T_0$, and an independently justified lower bound
$H_{\min}(K_{\rm acc}\mid E,T_0,F=1)\ge h_*$ holds. Suppose a reconciliation and verification transcript $C$ has at most $2^\lambda$ possible values and Alice's final string differs from Bob's with probability at most $\varepsilon_{\rm cor}$ conditional on completion. Two-universal privacy amplification with a uniformly random public seed yields an $\ell$-bit key satisfying
\begin{equation}\label{eq:appkey}
\ell\le h_*-\lambda-2\log_2\left(\frac{1}{2\varepsilon_{\rm PA}}\right),
\end{equation}
with trace-distance secrecy parameter at most $\varepsilon_{\rm PA}$ in the unsmoothed convention and total correctness failure at most $\varepsilon_{\rm cor}$.
\end{proposition}
\begin{proof}
Treating $C$ as classical public information, Proposition~\ref{prop:syndrome} (or its same-cardinality extension to general announced functions) reduces the lower bound on conditional min-entropy by at most $\lambda$. The quantum leftover hash lemma for a two-universal family and independent seed gives secrecy distance at most $\frac12\sqrt{2^{\ell-H_{\min}(K_{\rm acc}\mid E,T_0,C,F=1)}}$, which is no greater than $\varepsilon_{\rm PA}$ under~\eqref{eq:appkey}. The verification assumption independently bounds Alice--Bob disagreement by $\varepsilon_{\rm cor}$. If parameter estimation, authentication, or smoothing are added, their failure budgets and entropy penalties must be incorporated explicitly; they are not silently included in this proposition \cite{tom,leftover}.
\end{proof}

\begin{remark}[Nonadditivity]
A single accepted block has at most six raw bits, so a fixed finite secrecy target may leave no positive one-block extraction after disclosure and verification. A meaningful positive asymptotic rate requires a joint entropy lower bound on the actual multi-block adversarial state, potentially using a suitable entropy-accumulation or collective-attack reduction theorem. Neither block independence nor a tensor-power structure follows from the optical model as stated.
\end{remark}

\paragraph{Proof of optical-model insufficiency.}
\begin{proof}[Proof of Proposition~\ref{prop:nogo}]
The honest optical description specifies the action of Bob's filter on a designated signal mode and the detection statistics at Alice. It imposes no restriction on all other modes that can leave Bob's laboratory or on active incoming interrogation. Construct an extension of the described device that performs exactly the stipulated transformation on the designated signal mode but also makes the classical setting $K$ accessible to an adversarial output register, $|k\rangle\!\langle k|_{E_{\rm tag}}$. Tracing out this additional register leaves the stipulated Alice--Bob statistics unchanged, including the complete joint click pattern and graph acceptance. However Eve reads $E_{\rm tag}$ and guesses $K$ perfectly, so $P_{\rm guess}(K\mid E,M_0)=1$ and $H_{\min}(K\mid E,M_0)=0$. Thus no positive lower bound on adversarial min-entropy follows from the stipulated observables without extra device restrictions.
\end{proof}

The constructed tag is an abstract counterexample showing the incompleteness of the device specification, not a claim that every optical implementation literally transmits a classical side-channel register. A physical concern is active optical probing: Bob's choice of one of four polarization-transmission settings can influence the polarization of returned adversarial light, and multiple-photon probes can make nonorthogonal returned polarizations distinguishable with high confidence. Such probing must be bounded by an explicit input-light energy constraint, isolation, filtering, watchdog monitoring, trusted mode assumptions, or a source/device characterization; none of these restrictions is currently present in the ideal-model statement. The general relevance of Trojan-horse attacks to bidirectional modulating devices has been demonstrated in the literature \cite{trojan,gisin2006}. This gap is independent of the parity-graph topology.

\paragraph{Requirements for a full optical security proof.}
To turn the conditional result into an unconditional theorem, the model must: (i) specify the complete input and returned optical modes of Bob's device, including unwanted reflections, wavelength and temporal modes, and permitted photon number; (ii) define the class of Eve's joint forward--return attacks and the authenticated transcript; (iii) connect observable data, including test statistics beyond the accepted-block QBER, to an attack-uniform lower bound on smooth min-entropy of the raw settings; (iv) account for acceptance and public communications in finite size, with a joint many-block entropy argument; and (v) verify the correctness and secrecy failure budget. Until these are supplied, Theorem~\ref{thm:conditional} is a complete \emph{conditional postprocessing theorem} rather than a security proof of the physical QKD implementation.

\subsection{Passive-tap optical channel}\label{app:passiveproof}
\paragraph{Distinguishability of one-slot optical returns.}
\begin{proposition}[Passive-tap preselection entropy]\label{prop:prepassive}
Under the single-photon, independent, lossless passive beam-splitter model of Section~\ref{sec:passive}, even if Eve learns all preparation bits $a$, one obtains Eq.~\eqref{eq:preoptical}.
\end{proposition}
\begin{proof}
Fix a preparation bit $a$. For $b=a$, Bob returns one photon and Eve obtains the mixture in Eq.~\eqref{eq:evetapstates}; for $b\ne a$, she receives vacuum. These density operators commute and their trace distance is $\lambda$. Helstrom's binary minimum-error theorem gives $P_{\rm guess}(b\mid E,a)=(1+\lambda)/2$. Independent slots produce commuting product states. For each observed six-slot photon pattern the maximum a posteriori probability factorizes, so the optimal whole-string guessing probability is $[(1+\lambda)/2]^6$. Taking its negative base-two logarithm proves the claim. Quantum storage or coherent joint measurements do not improve guessing because the accessible states commute.
\end{proof}

\paragraph{Accepted-branch distribution and optimal guessing.}
\begin{proposition}[Exact min-entropy after public filtering]\label{prop:exactpassive}
With the same channel and the public two-check syndrome $S=HK$, Eve's optimal whole-block guessing probability conditional on acceptance is Eq.~\eqref{eq:exactguess}. In particular it depends on the selected tap pattern $t$ and must not be evaluated by treating acceptance as independent of Eve.
\end{proposition}
\begin{proof}
Since $a$ is public to Eve by the conservative model assumption, relabel $d=K\xor a$. Bob transmits exactly if $d_i=0$. Eve's binary detection record $t_i=1$ has probability $\lambda$ when $d_i=0$ and zero when $d_i=1$. Consequently
\begin{equation}\label{eq:jointpassive}
 \Pr(d,t\mid a)=2^{-6}\,\mathbf1_{t\wedge d=0}
 \lambda^{|t|}(1-\lambda)^{6-|t|-|d|}.
\end{equation}
Alice loses precisely tapped photons; hence her bit inference error equals $t$. In particular $F=1\Leftrightarrow Ht=0$. The observed public syndrome is $Hd=s$ up to the known additive offset $Ha$. Given $(t,s,F=1)$, Eve's optimal prediction of $d$ is the largest joint weight among vectors with $Hd=s$ and $t\wedge d=0$. Summing this maximum over the mutually exclusive classical observations $(t,s)$ and dividing by $p_{\rm acc}$ proves Eq.~\eqref{eq:exactguess}. The optimality is quantum-valid within this channel because all conditional Eve states are diagonal in the vacuum--one-photon number basis, and a photon-number measurement loses no information. The accepted branch changes Eve's joint distribution; the proof takes this dependence into account exactly.
\end{proof}

\paragraph{Acceptance probability and reconciliation entropy.}
\begin{proposition}[Exact restricted-channel key-rate quantities]\label{prop:taprate}
Let $p_{\rm acc}=\sum_{d,t}\Pr(d,t)\mathbf1_{Ht=0}$, and define the accepted conditional mass function
\begin{equation}
 \pi(d,t)=\frac{\Pr(d,t)\mathbf1_{Ht=0}}{p_{\rm acc}}.
\end{equation}
Then $p_{\rm acc}=[1-2e(1-e)]^3+[2e(1-e)]^3$ with $e=\lambda/2$, and the entropies in Eq.~\eqref{eq:passiverate} follow exactly by grouping $\pi(d,t)$ according to Eve's pair $(t,Hd)$ and Alice's pair $(d\xor t,Hd)$, respectively. The per-block asymptotic entropy difference is an achievable one-way private randomness rate in the stated iid classical--quantum model, provided authenticated reconciliation and privacy amplification attain their asymptotic rates.
\end{proposition}
\begin{proof}
For a uniform $b$ and fixed $a$, the tap indicator is Bernoulli with success probability $\lambda/2=e$, independently at each slot. The three vertex-error parities are therefore independent Bernoulli with parameter $q=2e(1-e)$. The graph accepts precisely when their values are constant, giving $p_{\rm acc}=(1-q)^3+q^3$. The public syndrome constrains $Hd$ and Eve's observation includes $t$. After acceptance the normalized probabilities are exactly $\pi(d,t)$. Shannon conditional entropy is determined uniquely by partitioning the finite outcomes according to the stated observations; the corresponding sums are implemented by \texttt{code/verify\_passive\_security.py} without approximation. Since $K$ is uniform and Eve's conditional states commute in this channel, her quantum conditional von Neumann entropy equals the classical entropy of $d$ given $(t,Hd,F=1)$. The standard iid one-way privacy amplification/reconciliation theorem gives Eq.~\eqref{eq:passiverate}. This asymptotic conclusion cannot be extended without proof to adaptive or coherent attacks outside the stipulated channel.
\end{proof}

\begin{remark}[Conservative composable finite-key extraction]
For $m$ independent accepted blocks, the unsmoothed conditional min-entropy is additive under the product assumption, with $m$ times the exact per-block value from Proposition~\ref{prop:exactpassive}. If an authenticated error-correction and verification transcript on those $m$ blocks has at most $2^L$ possible values and correctness failure at most $\varepsilon_{\rm cor}$, then the quantum leftover hash lemma yields secrecy error at most $\varepsilon_{\rm PA}$ for
\begin{equation}
 \ell\le m H_{\min}(K\mid E,a,S,F=1)-L-2\log_2\frac1{2\varepsilon_{\rm PA}}.
\end{equation}
The independent accepted-block assumption, a valid actual leakage bound $L$, and authenticated announcements are essential. The asymptotic Shannon cost is not itself a finite-length code or an explicit leakage guarantee.
\end{remark}

\paragraph{Limits of optical certification.}
The passive channel in this appendix is a \emph{restricted attack model}. Monitoring a few port transmissions does not establish that arbitrary Eve states satisfy Eq.~\eqref{eq:evetapstates}. For a device with unrestricted injected modes, attach to the stipulated honest output an orthogonal setting-dependent state $|k\rangle\langle k|$ accessible in another mode. This extension preserves all honest click probabilities but sets $H_{\min}(K\mid E)=0$. To exclude it, a model must specify and characterize input-light constraints, output-mode isolation, source photon-number tails, detector response, and admissible joint attacks. Characterization of those quantities is a necessary experimental program, but no measurement-dependent attack-uniform entropy theorem is supplied here. Therefore the present proof is complete within its assumptions but does not close general optical QKD security.

\clearpage
\begingroup\small\setlength{\parskip}{0pt}
\endgroup
\end{document}